\documentclass[preprint]{aastex62}

\graphicspath{{./}{}}

\shorttitle{Predicting UV Emission}
\shortauthors{Melbourne et al.}

\usepackage{subfigure}
\usepackage{color}
\usepackage{longtable}
\usepackage{amsmath}
\usepackage{hyperref}
\usepackage[pass]{geometry}
\usepackage{comment}

\newcommand{\Lya}{\mbox{Ly$\alpha$}}
\newcommand{\Ha}{\mbox{H$\alpha$}}

\newcommand{\HI}{H\,\textsc{i}}
\newcommand{\cahk}{\ion{Ca}{2} H\&K}
\newcommand{\ca}{\ion{Ca}{2}}
\newcommand{\MgII}{\ion{Mg}{2}}
\newcommand{\SiII}{\ion{Si}{2}}
\newcommand{\SiIII}{\ion{Si}{3}}

\newcommand{\CII}{\ion{C}{2}}

\newcommand{\rhk}{R$'_{HK}$}

\newcommand{\Msun}{M_{\odot}}
\newcommand{\Lhabol}{log$_{10}$ L$_{H\alpha}$/L$_{bol}$}

\newcommand{\newpara}
    {
    \vskip 0.1in
    \noindent
    }

\begin{document}

\title{Estimating the Ultraviolet Emission of M dwarfs with Exoplanets from Ca II and H$\alpha$}

\author[0000-0002-8423-6904]{Katherine Melbourne}
\affiliation{Astronomy Department, Yale University, 52 Hillhouse Ave, New Haven, CT 06511, USA}
\affiliation{Ball Aerospace \& Technologies Corp., 1600 Commerce Street, Boulder, CO 80021, USA}
\email{melbournekatherine@gmail.com}

\author[0000-0002-1176-3391]{Allison Youngblood}
\affiliation{NASA Goddard Space Flight Center, Greenbelt, MD 20771, USA}
\affiliation{Laboratory for Atmospheric and Space Physics, University of Colorado, 600 UCB, Boulder, CO 80309, USA}

\author[0000-0002-1002-3674]{Kevin France}
\affiliation{Laboratory for Atmospheric and Space Physics, University of Colorado, 600 UCB, Boulder, CO 80309, USA}

\author{C. S. Froning}
\affiliation{McDonald Observatory, University of Texas at Austin, Austin, TX 78712, USA}

\author[0000-0002-4489-0135]{J. Sebastian Pineda}
\affiliation{Laboratory for Atmospheric and Space Physics, University of Colorado, 600 UCB, Boulder, CO 80309, USA}

\author[0000-0002-7260-5821]{Evgenya L. Shkolnik}
\affiliation{School of Earth and Space Exploration, Arizona State University, 781 Terrace Mall, Tempe, AZ 85287, USA}

\author[0000-0001-9667-9449]{David J. Wilson}
\affiliation{McDonald Observatory, University of Texas at Austin, Austin, TX 78712, USA}

\author[0000-0002-4998-0893]{Brian E. Wood}
\affiliation{Naval Research Laboratory, Space Science Division, Washington, DC 20375, USA}

\author[0000-0002-6163-3472]{Sarbani Basu}
\affiliation{Astronomy Department, Yale University, 52 Hillhouse Ave, New Haven, CT 06511, USA}

\author[0000-0002-2989-3725]{Aki Roberge}
\affiliation{Exoplanets and Stellar Astrophysics Lab, NASA Goddard Space Flight Center, Greenbelt, MD 20771, USA}

\author{Joshua E. Schlieder}
\affiliation{Exoplanets and Stellar Astrophysics Lab, NASA Goddard Space Flight Center, Greenbelt, MD 20771, USA}

\author[0000-0001-9207-0564]{P. Wilson Cauley}
\affiliation{Laboratory for Atmospheric and Space Physics, University of Colorado, 600 UCB, Boulder, CO 80309, USA}

\author[0000-0001-5646-6668]{R. O. Parke Loyd}
\affiliation{School of Earth and Space Exploration, Arizona State University, Tempe, AZ 85287, USA}

\author{Elisabeth R. Newton}
\affiliation{Departmant of Physics and Astronomy, Dartmoth College, Hanover, NH 03755, USA}

\author{Adam Schneider}
\affiliation{School of Earth and Space Exploration, Arizona State University, Tempe, AZ 85287, USA}

\author{Nicole Arulanantham}
\affiliation{Laboratory for Atmospheric and Space Physics, University of Colorado, 600 UCB, Boulder, CO 80309, USA}

\author[0000-0002-3321-4924]{Zachory Berta-Thompson}
\affiliation{Department of Astrophysical \& Planetary Sciences, 391 UCB 2000, Boulder, CO 80309, USA}

\author{Alexander Brown}
\affiliation{Center for Astrophysics and Space Astronomy, University of Colorado, 389 UCB, Boulder, Colorado 80309, USA}

\author{Andrea P. Buccino}
\affiliation{Instituto de Astronomia y Fisica del Espacio (CONICET--UBA), Buenos Aires, Argentina}

\author{Eliza Kempton}
\affiliation{University of Maryland, College Park, MD 20742}

\author{Jeffrey L. Linsky}
\affiliation{Joint Institute for Laboratory Astrophysics, University of Colorado and NIST, Boulder, CO 80309-0440, USA}

\author[0000-0002-9632-9382]{Sarah E. Logsdon}
\affiliation{Exoplanets and Stellar Astrophysics Lab, NASA Goddard Space Flight Center, Greenbelt, MD 20771, USA}
\affiliation{NSF's National Optical-Infrared Astronomy Research Laboratory, 950 North Cherry Ave., Tucson, AZ 85719, USA}

\author{Pablo Mauas}
\affiliation{Instituto de Astronomia y Fisica del Espacio (CONICET--UBA), Buenos Aires, Argentina}

\author[0000-0001-9573-4928]{Isabella Pagano}
\affiliation{National Institute of Astrophysics, Cantania Astrophysical Observatory, Via S. Sofia 78, 95123, Catania, Italy}

\author[0000-0002-1046-025X]{Sarah Peacock}
\affiliation{Lunar and Planetary Laboratory, University of Arizona, 1629 E University Blvd, Tucson, AZ 8572, USA}

\author[0000-0003-3786-3486]{Seth Redfield}
\affiliation{Astronomy Department and Van Vleck Observatory, Wesleyan University, Middletown, CT 06459, USA}

\author[0000-0003-1620-7658]{Sarah Rugheimer}
\affiliation{University of Oxford, Atmospheric, Oceanic, and Planetary Physics Department, Clarendon Laboratory, Sherrington Rd, Oxford OX1 3PU, UK}

\author{P. Christian Schneider}
\affiliation{Hamburger Sternwarte, Gojenbergsweg 112, 21029 Hamburg, Germany}

\author[0000-0002-1912-3057]{D. J. Teal}
\affiliation{NASA Goddard Space Flight Center, Greenbelt, MD 20771, USA}
\affiliation{University of Maryland, College Park, MD 20742}

\author[0000-0002-9607-560X]{Feng Tian}
\affiliation{Macau University of Science and Technology, Avenida Wai Long, Taipa, Macau}

\author{Dennis Tilipman}
\affiliation{Department of Astrophysical \& Planetary Sciences, 391 UCB 2000, Boulder, CO 80309, USA}

\author{Mariela Vieytes}
\affiliation{Instituto de Astronomia y Fisica del Espacio (CONICET--UBA), Buenos Aires, Argentina}
\affiliation{National University of Tres de Febrero, 2736, AHF, Av. Gral. Mosconi, B1674 Sáenz Peña, Buenos Aires, Argentina}

\begin{abstract}

M dwarf stars are excellent candidates around which to search for exoplanets, including temperate, Earth-sized planets. To evaluate the photochemistry of the planetary atmosphere, it is essential to characterize the UV spectral energy distribution of the planet's host star. This wavelength regime is important because molecules in the planetary atmosphere such as oxygen and ozone have highly wavelength dependent absorption cross sections that peak in the UV (900-3200 \AA{}). We seek to provide a broadly applicable method of estimating the UV emission of an M dwarf, without direct UV data, by identifying a relationship between non-contemporaneous optical and UV observations. Our work uses the largest sample of M dwarf star far- and near-UV observations yet assembled. We evaluate three commonly-observed optical chromospheric activity indices -- H$\alpha$ equivalent widths and \Lhabol, and the Mount Wilson \cahk{} S and \rhk{} indices -- using optical spectra from the HARPS, UVES, and HIRES archives and new HIRES spectra. Archival and new Hubble Space Telescope COS and STIS spectra are used to measure line fluxes for the brightest chromospheric and transition region emission lines between 1200-2800 \AA{}. Our results show a correlation between UV emission line luminosity normalized to the stellar bolometric luminosity and Ca II \rhk{} with standard deviations of 0.31-0.61 dex (factors of $\sim$2-4) about the best-fit lines. We also find correlations between normalized UV line luminosity and H$\alpha$ \Lhabol{} and the S index. These relationships allow one to estimate the average UV emission from M0 to M9 dwarfs when UV data are not available.

\end{abstract}


\section{Introduction}\label{sec:Introduction}

M dwarfs (M$_{\star}$ $\lesssim 0.6 \Msun$) are excellent candidates for ongoing exoplanet detection and characterization efforts \citep{Shields2017TheStars, Tarter2007SpecialStars}. These are the most abundant stars in the stellar neighborhood (RECONS Survey\footnote{\url{www.recons.org}}), making up $\sim$75 percent of the Galaxy's total stellar population \citep{Bochanski2010THEFIELD, Reid1997Low-MassFunction}. M dwarfs are known to harbor larger populations of small and terrestrial planets relative to solar-type stars \citep{Batalha2013PLANETARYDATA, Dressing2015THESENSITIVITY}. The signals from planets orbiting M dwarfs are larger and easier to identify for both the transit and radial velocity methods, the two most commonly-used techniques for finding exoplanets. In addition, the long lifetime of an M dwarf (on the order of $10^{12}$ of years \citep{Laughlin1997THESEQUENCE} for the lowest mass stars) provides ample time for planets to form and life to evolve. Despite concerns about extreme ultraviolet (EUV) emission, frequent energetic flares, coronal mass ejections, and possible tidal locking, which could alter the atmosphere of an exoplanet orbiting an M dwarf (see \cite{Shields2017TheStars} and references within), these smaller, dimmer stars remain a priority in current exoplanet detection efforts due to their many advantages.

The high-energy radiative environment around M dwarfs can significantly impact the upper atmospheric temperature and chemistry of the exoplanets orbiting them \citep{Segura2005BiosignaturesDwarfs, Lammer2007CoronalZones, Miguel2015The436b, Rugheimer2015EFFECTSTARS, Arney2017PaleExoplanets}. Quantifying the $\sim$1200-1700 \AA{} (far-UV hereafter) and 1700-3200 \AA{} (near-UV) wavelength ranges of ultraviolet emission of an M dwarf is essential to assessing any photochemical byproducts (e.g., hazes, biosignatures) in the atmospheres of its exoplanets \citep{Shields2017TheStars, Meadows2018ExoplanetEnvironment}. In addition, knowledge of an M dwarf's far-UV (FUV) and near-UV (NUV) radiation environment allows for the estimation of EUV emission \citep{Linsky2013THESTARS, France2018Far-UltravioletStars, Peacock2019PredictingSystem, Sanz-Forcada2011AstronomyEvaporation, Chadney2015XUV-drivenStars}. This is significant because EUV observations cannot be taken directly for stars other than the Sun; there is no current dedicated mission yet that observes 170--900 \AA{}, and interstellar medium (ISM) attenuation makes observations between $\sim$500-911 \AA{} unfeasible for many stars. 

M dwarfs have bolometric luminosities ranging from $\sim$10$^{-4}$ -- 10$^{-1}$ L$_{\odot}$, moving the location of cool exoplanets (including liquid-water habitable zone (HZ) planets) 5 to 100 times closer around an M dwarf than the comparable orbital separation from the Sun. Combining this with the relatively high UV fluxes of M dwarfs, the incident FUV fluxes are generally higher for cool planets ($T_{equil} \lesssim 1000$ K) orbiting M dwarfs than they would be around stars of other spectral classes. Photochemical processes depend on the far- to near- UV flux ratio \citep{France2013THESTARS, Loyd2016ThePlanets} and can influence the formation of hazes \citep{Morley2013,Morley2015,Crossfield2017,LibbyRoberts2020} as well as the abundances of key biosignatures (O$_2$, O$_3$, and CH$_4$) and habitability indicators (CO$_2$ and H$_2$O) in exoplanet atmospheres around M dwarfs \citep{Hu2012PhotochemistryCases, 2014ApJ...792...90D, Tian2013HighPlanets, 2014ApJ...785L..20W, Gao2015STABILITYEXOPLANETS, 2015ApJ...812..137H, Luger2015ExtremeDwarfs}. For example, Lyman $\alpha$ (Ly$\alpha = 1215.67$), the brightest UV emission line in an M dwarf spectrum, has been shown to significantly alter the H$_2$O mixing ratios within the atmosphere of a mini-Neptune orbiting its M dwarf host \citep{Miguel2015The436b}. As H$_2$O and CO$_2$ in a terrestrial planet atmosphere absorb Ly$\alpha$ radiation from the M dwarf, they dissociate, creating free H and O atoms and OH and CO molecules, leading to catalytic cycles that can produce false-positive and false-negative biosignatures \citep{2015ApJ...812..137H, Miguel2015The436b, Rugheimer2015EFFECTSTARSb}. UV radiation also leads to the formation of organic hazes in rocky planet atmospheres \citep{Arney2017PaleExoplanets} as well as hazes in gaseous planets. Haze strongly affects an exoplanet's spectral features as well as habitability \citep{Horst2018ExploringComposition,Arney2018OrganicAtmospheres}, and an accurate UV spectrum is critical to haze formation modeling. 

To analyze exoplanet atmospheres and identify potential false-positive biosignatures, it is essential to characterize the spectral energy distributions (SEDs) of M dwarf hosts, as the cross-sections of important molecules and atoms are highly wavelength-dependent and peak in the UV \citep{Hu2012PhotochemistryCases}. Most stellar models do not extend beyond the photosphere \citep{2013A&A...553A...6H,Allard2011ModelsExoplanets,Hauschildt1998TheK}, thus excluding the primary regions of UV emission. Some progress has been made in including chromospheres, transition region, and coronae for models of individual stars \citep{Fontenla2016semi, Peacock2019PredictingSystem}, but those methods are not yet broadly applicable. As a result, SEDs must be directly observed to measure the FUV and NUV activity of a specific star. This requires dedicated space-based UV telescopes and underscores the importance of current observations by the Hubble Space Telescope (HST) \citep{France2016TheOverview,Guinan2016LivingStar,Loyd2018HAZMAT.Ultraviolet,Ribas2017TheCentauri}. However, there may be a gap in observing capability for UV characterization after HST stops UV observations in the coming years. As the Transiting Exoplanet Survey Satellite (TESS) has completed its primary mission, the Characterizing Exoplanets Satellite (CHEOPS) has completed its commissioning, the James Webb Space Telescope (JWST) launch approaches, and the extremely large telescopes (ELTs) prepare to begin operations in the late 2020s, a lapse in UV spectral data would limit characterization of exoplanets discovered with these instruments. Consequently, it is crucial to identify an alternative approach to estimating the UV emission of M dwarfs from optical data measured with ground-based observatories.

Past research has demonstrated the effectiveness of using the \cahk{} resonance lines (3969 \AA{}, 3934 \AA{}) as indicators for stellar activity (e.g., \citealt{1963ApJ...138..832W,Cincunegui2007AstrophysicsCentauri,Walkowicz2009TRACERSDWARFS}). \cahk{} lines appear against the continuum as a superposition of broad Ca$^+$ absorption ($>1$\AA{}) from the cool upper photosphere and lower chromosphere as well as narrow Ca$^+$ emission ($<-$0.5\AA{}) from the hot upper chromosphere \citep{Fontenla2016semi}. \cahk{} emission has been studied in detail since the Mount Wilson observing program in the 1960s \citep{Wilson1968FLUXK-LINES} through two stellar activity indices: the S index that includes both chromospheric and photospheric emissions, and \rhk{}, a transformation of the S index that is normalized to the bolometric flux. This enables comparisons between different spectral types by excluding the contribution of photospheric emission to the measured S index. The S index is only a normalized measure of the line core emission and does not provide a value for the absolute energy emitted in the line, so it is essential to determine \rhk{} as well. The program at Mount Wilson defined S index and \rhk{} exclusively using observations of F, G, and K stars, and until recently, the color indices used to calculate these parameters had not been well-calibrated for M dwarfs. \citealt{Cincunegui2007AstrophysicsStars, SuarezMascareno2015} used flux calibrated observations to extend to M dwarfs the color correction factors needed to correct the S index for spectral type effects. \citealt{Astudillo-Defru2016MagneticRHK} reexamined both the S index and \rhk{} \cahk{} activity tracers for FGK stars and identified an improved method of accurately calculating S index and \rhk{} for M dwarfs, which we follow in this paper.

The \Ha{} (6562.8 \AA{}) equivalent width (EW) and \Ha{} luminosity normalized to the stellar bolometric luminosity (\Lhabol) were selected for this work as they are also a commonly used indicator for stellar activity and measurements are widely available in the literature \citep{Reid1995TheKinematics, Gizis2002TheHistory, West2011THEDATA, 2014ApJ...795..161D, Gaidos2014TrumpetingLife, Alonso-Floriano2015AstrophysicsCAFOS, Newton2017TheRotation}. \Ha{} traces the top of the chromosphere \citep{Mauas1993ATMOSPHERICLEO,Mauas1996AtmosphericLeonis,Leenaarts2012THECHROMOSPHERE}. All M dwarfs appear to emit significant UV radiation \citep{France2016TheOverview}, but they are still categorized in the literature as being either ``active" or ``inactive" dependent on each star's \Ha{} emission. For stars in the ``inactive" regime (\Ha{} EW $>$ -1 \AA, i.e. in absorption), including a significant fraction of the stellar sample used for this work, \Ha{} has been shown to be non-monotonic with stellar activity \citep{Cram1985FORMATIONSTARS, Stauffer1986CHROMOSPHERIC1} and therefore may not be a precise tracer of UV activity. However, \cite{Newton2017TheRotation} showed \Lhabol{} to be significantly correlated with stellar rotation period.

The correlation between M dwarf chromospheric optical and UV emission has been demonstrated previously (e.g., \Ha{}-CIV, \Ha{}-MgII, \ca{} K-MgII, and \ca{} K with several Balmer lines) \citep{Hawley1991TheLeonis, Hawley2003TRANSITIONSTARS, Walkowicz2008TracersDwarfs, Youngblood2017TheDwarfs}. \cite{Youngblood2017TheDwarfs} identified the relationships between \ca{} and nine UV spectral lines for 15 M dwarfs from 1200-2800 \AA. In this paper, we expand the sample to 69 M dwarfs and increase the range of activity levels, spectral types, and ages of the stellar sample.

This paper is organized as follows: Section \ref{sec:ObservationsReductions} details the M dwarf target selection and the observations used for the analysis in this work. Section \ref{sec:HaEquivalentWidths} and Section \ref{sec:CaIIIndices} discuss how measurements of each optical activity index were found with spectral analysis of the \Ha{} and \cahk{} lines, respectively. Section 2 also includes a description of how the UV emission line fluxes were determined. Section \ref{sec:UVOpticalRelations} describes the correlations found between each optical activity indicator and each UV emission line. We conclude with a discussion of these results in Section \ref{sec:Discussion} and a summary of the findings in Section \ref{sec:Summary}.

\section{Observations and Reductions}\label{sec:ObservationsReductions}

\begin{figure}
\centering
\includegraphics[width=14cm]{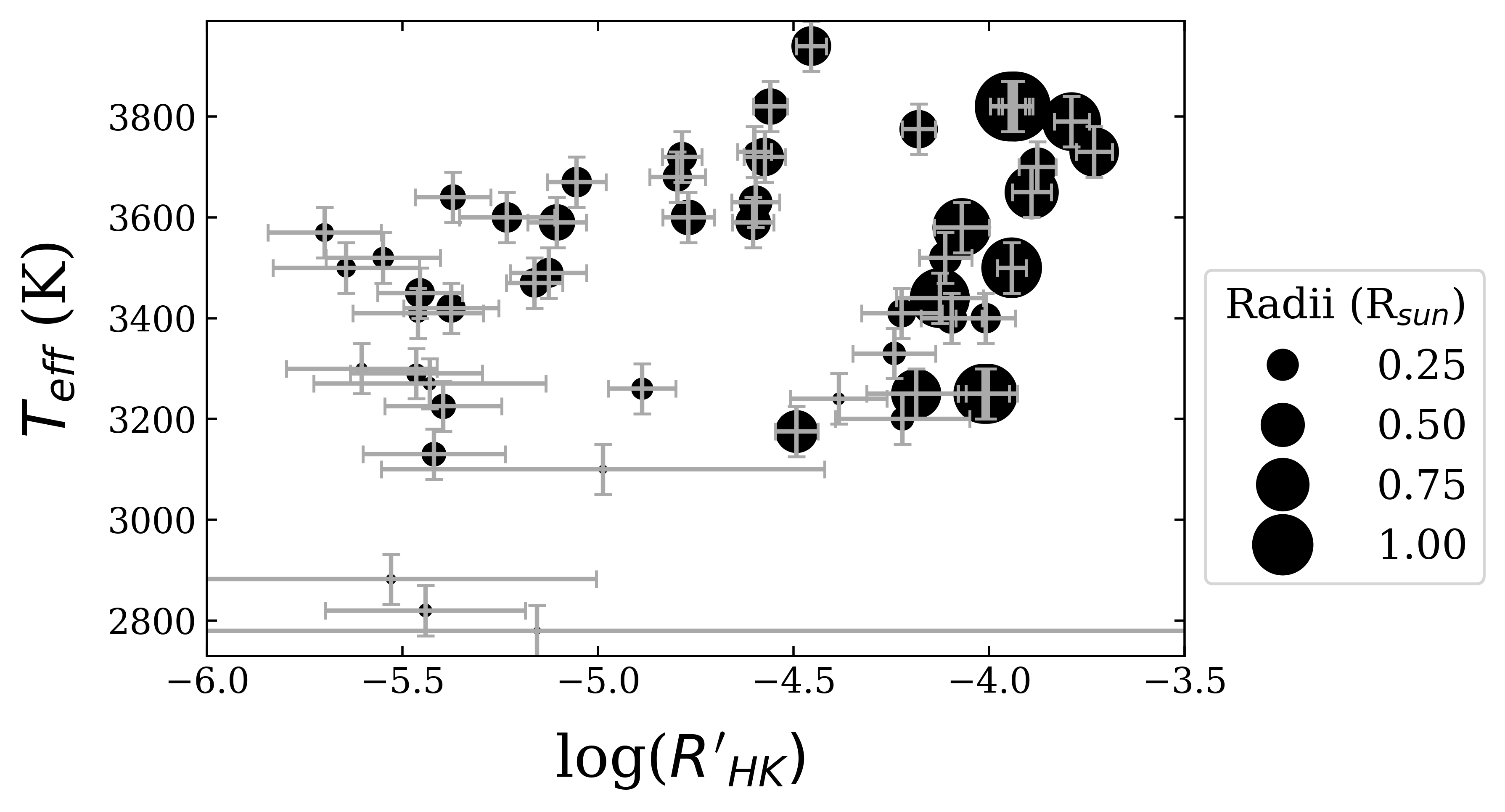}
\caption{The M dwarf sample studied in this work spans a range of $\sim$1000 K in effective temperature, with an assumed uncertainty of $\pm$50 K on each value, and $\sim$3 dex in activity levels as represented by the log \rhk{}~values (see Section~\ref{sec:CaIIIndices}). The relative radius of each star is demonstrated by the size of the points, with a range of 0.11 to 1.07 R$_{\odot}$. Young, early M dwarfs are included in this sample that have radii around 1 R$_{\odot}$ (see Table \ref{table:machinereadabletable} for further clarification). This plot shows the 45 stars (of 69 total targets) for which an \rhk{} value is reported in this work.}
\label{temp_vs_rhk}
\end{figure}

The M dwarfs analyzed in this work were chosen based on the availability of UV spectra from HST and Ca II and/or H$\alpha$~spectra from high-resolution optical spectrographs like HARPS, HIRES, and UVES. Many of the UV spectra came from recent HST guest observer programs, including the MUSCLES Treasury Survey (GO-13650; \citealt{France2016TheOverview, Youngblood2016TheExoplanets, Loyd2016ThePlanets}), the Mega-MUSCLES Treasury Survey (GO-15071; \citealt{2019ApJ...871L..26F}), the Far Ultraviolet M-dwarf Evolution Survey (FUMES) (GO-14640; Pineda et al. 2020 Under Review, Youngblood et al. 2020 in prep.), the Habitable Zones and M dwarf Activity across Time (HAZMAT) survey (GO-14784; \citealt{Loyd2018HAZMAT.Ultraviolet}), and the M Dwarf Stellar Wind survey (GO-15326; Wood et al. in preparation). Our targets have range in spectral type from M0 to M9, and all but one target observed are within $d < 60$ pc of Earth. The sample covers a wide range of ages ($\sim$0.01 to $\sim$10 Gyr): several stars in the TW Hya association are likely the youngest stars ($\sim$10 Myr; \citealt{Weinberger2012DISTANCEPARALLAXES}) and Barnard's Star and Kapteyn's Star are likely the oldest stars ($\sim$10 Gyr; \citealt{2018Natur.563..365R,Kotoneva2005AstronomyStar,Wylie-DeBoer2010EVIDENCEGROUP}). Figure \ref{temp_vs_rhk} provides a visual of the range of radii, effective temperatures, and \rhk{}; other details about each target may be found in Tables \ref{table:targlist} and \ref{table:machinereadabletable}. Ages were gathered from the literature, where they were established through membership to young moving groups or clusters or galactic kinematics. In the absence of any age information, targets are assumed to be field age ($\sim$5 Gyr).

\begin{deluxetable}{c|c|c|c}
\tabletypesize{\scriptsize}
\movetabledown=20mm
\setlength{\tabcolsep}{3pt}
\tablewidth{0pt}
\tablecaption{The M dwarf sample \label{table:targlist}} 
\tablehead{\colhead{\textbf{Star}} & 
                  \colhead{\textbf{Star}} &
                  \colhead{\textbf{Star}} &
                  \colhead{\textbf{Star}}
                  }
\startdata
\textbf{GO 13650 (MUSCLES)$^1$}	&	GJ173	&	\textbf{AR 10638 (STARCat)$^5$}	&	\textbf{GO 9090$^{11}$}		\\
GJ176	&	GJ3290	&	GJ388	&	GJ285		\\
GJ667C	&	LP5-282	&	GJ803	&	GJ644C		\\
GJ581	&	2MASSJ04223953+1816097	&	GJ873	&			\\
GJ1214	&	2MASSJ04184702+1321585	&	GJ551	&	\textbf{GO 15071 (Mega-}		\\
GJ832	&	2MASSJ02125819-5851182	&		&	\textbf{MUSCLES)$^{12}$}		\\
GJ876	&	GJ3997	&	\textbf{GO 14462$^6$}	&	TRAPPIST-1		\\
GJ436	&	2MASSJ22463471-7353504	&	GJ1132	&	GJ676A		\\
GJ628	&		&		&	GJ15A		\\
GJ887	&	\textbf{GO 14640 (FUMES)$^3$}	&	\textbf{GO 14767 (PanCET)$^7$}	&	GJ649		\\
GJ1061	&	GJ4334	&	GJ3470	&	GJ163		\\
HD173739	&	GJ49	&		&	GJ849		\\
	&	GSC07501-00987	&	\textbf{GO 12361$^8$}	&	GJ674		\\
\textbf{GO 14784 (HAZMAT)$^2$}	&	LP247-13	&	TWA13A	&	GJ699		\\
GSC8056-0482	&	G80-21	&	TWA13B	&	LHS2686		\\
2MASSJ02543316-5108313	&	CD-571054	&		&	GJ729		\\
2MASSJ02001277-0840516	&	CD-352722	&	\textbf{GO 11616$^9$}	&			\\
G75-55	&	GJ410	&	TWA7	&	\textbf{GO 15326$^{13}$}		\\
2MASSJ22025453-6440441	&	LP55-41	&		&	GJ273		\\
2MASSJ00240899-6211042	&	G249-11	&	\textbf{GO 12011$^{10}$}	&	GJ205		\\
2MASSJ01521830-5950168	&		&	LHS2065	&	GJ588		\\
2MASSJ03315564-4359135	&	\textbf{GO 13020 (Living with}	&	LHS3003	&	GJ338A		\\
2MASSJ23261069-7323498	&	\textbf{a Red Dwarf)$^4$}	&		&			\\
2MASSJ23285763-6802338	&	GJ191	&		&	\textbf{GO 15190$^{14}$}		\\
2MASSJ00393579-3816584	&		&		&	GJ411		\\
\enddata
\tablecomments{(1) \citealt{France2016TheOverview}, (2) \citealt{Loyd2018HAZMAT.Ultraviolet}, (3) Pineda et al. in prep., (4) \citealt{Guinan2016LivingStar}, (5) \citealt{Ayres2010StarCAT:STARS}, (6) \citealt{Waalkes2019LyIOPscience}, (7) \citealt{Sing2019TheIOPscience}, (8) PI: Brown, (9) \citealt{2012ApJ...756..171F}, (10) PI: Osten, (11) \citealt{Hawley2003TransitionStarsb}, (12) \citealt{2019ApJ...871L..26F}, (13) Wood et al. in prep., (14) Youngblood et al. in prep.}
\end{deluxetable}

\begin{deluxetable}{cccl}
\tabletypesize{\scriptsize}
\movetabledown=20mm
\setlength{\tabcolsep}{3pt}
\tablewidth{0pt}
\tablecaption{M dwarf parameters, optical activity indicators, and UV luminosities. (Machine readable version available online) \label{table:machinereadabletable}} 
\tablehead{\colhead{\textbf{Column}} & 
                  \colhead{\textbf{Format}} &
                  \colhead{\textbf{Units}} &
                  \colhead{\textbf{Description}}
                  }
\startdata
1	&	str	&	---	&	Target name				\\
2	&	int	&	---	&	Identifier used in the paper				\\
3	&	flt	&	pc	&	Stellar distance				\\
4	&	str	&	---	&	ADS bibliography code reference for distance				\\
5	&	str	&	---	&	Stellar spectral type				\\
6	&	str	&	---	&	ADS bibliography code reference for spectral type				\\
7	&	flt	&	K	&	Stellar effective temperature				\\
8	&	str	&	---	&	ADS bibliography code reference for effective temperature				\\
9	&	flt	&	R$_{\odot}$	&	Stellar radius				\\
10	&	str	&	---	&	ADS bibliography code reference for radius				\\
11	&	flt	&	mag	&	V magnitude				\\
12	&	str	&	---	&	ADS bibliography code reference for V			\\
13	&	flt	&	mag	&	K magnitude				\\
14	&	str	&	---	&	ADS bibliography code reference for K				\\
15	&	flt	&	Gyr	&	Stellar age				\\
16	&	str	&	---	&	ADS bibliography code reference for age				\\
17	&	flt	&	0.1nm	&	\Ha{} equivalent width				\\
18	&	flt	&	0.1nm	&	Uncertainty in \Ha{} equivalent width				\\
19	&	str	&	---	&	ADS bibliography code reference for \Ha{}				\\
20	&	flt	&	---	&	S index				\\
21	&	flt	&	---	&	Uncertainty in S-index				\\
22	&	flt	&	---	&	log$_{10}$ R$\prime$HK				\\
23	&	flt	&	---	&	Uncertainty in log$_{10}$ R$^{\prime}$HK				\\
24	&	str	&	---	&	ADS bibliography code reference for S-index and log$_{10}$ R$\prime$HK				\\
25	&	str	&	---	&	Instruments used for optical activity indicators \\
    &           &        & (H = HARPS, K = HIRES, U = UVES, M = MIKE, \\
    &           &        & E = ELODIE, S = ESPaDOnS)		\\
26	&	flt	&	---	&	Number of \Ha{} spectra used				\\
27	&	flt	&	---	&	Number of Ca II spectra used				\\
28	&	flt	&	erg/s	&	Si III Luminosity				\\
29	&	flt	&	erg/s	&	Si III Luminosity uncertainty				\\
30	&	str	&	---	&	ADS bibliography code reference for Si III				\\
31	&	flt	&	erg/s	&	\Lya{} Luminosity				\\
32	&	flt	&	erg/s	&	\Lya{} Luminosity uncertainty				\\
33	&	str	&	---	&	ADS bibliography code reference for \Lya{}				\\
34	&	flt	&	erg/s	&	Si II Luminosity				\\
35	&	flt	&	erg/s	&	Si II Luminosity uncertainty				\\
36	&	str	&	---	&	ADS bibliography code reference for Si II				\\
37	&	flt	&	erg/s	&	C II (1335 \AA) Luminosity 				\\
38	&	flt	&	erg/s	&	C II (1335 \AA) Luminosity uncertainty				\\
39	&	str	&	---	&	ADS bibliography code reference for C II				\\
40	&	flt	&	erg/s	&	Mg II Luminosity				\\
41	&	flt	&	erg/s	&	Mg II Luminosity uncertainty				\\
42	&	str	&	---	&	ADS bibliography code reference for Mg II				\\
43	&	flt	&	erg/s	&	Si IV Luminosity				\\
44	&	flt	&	erg/s	&	Si IV Luminosity uncertainty				\\
45	&	str	&	---	&	ADS bibliography code reference for Si IV				\\
46	&	flt	&	erg/s	&	He II Luminosity				\\
47	&	flt	&	erg/s	&	He II Luminosity uncertainty			\\
48	&	str	&	---	&	ADS bibliography code reference for He II				\\
49	&	flt	&	erg/s	&	C IV Luminosity				\\
50	&	flt	&	erg/s	&	C IV Luminosity uncertainty				\\
51	&	str	&	---	&	ADS bibliography code reference for C IV				\\
52	&	flt	&	erg/s	&	N V Luminosity				\\
53	&	flt	&	erg/s	&	N V Luminosity uncertainty				\\
54	&	str	&	---	&	ADS bibliography code reference for N V				\\
\enddata
\end{deluxetable}

\subsection{Optical Data}\label{subsec:Optical}

For the optical spectra, we used public archival data from three ground-based observatories, the High Accuracy Radial velocity Planet Searcher (HARPS) on the ESO 3.6-meter telescope \citep{Mayor2003SettingHARPS}, the Ultraviolet and Echelle Spectrograph (UVES) on the Very Large Telescope \citep{Dekker2000DesignObservatory}, and the High Resolution Echelle Spectrometer (HIRES) on the Keck I Telescope \citep{Vogt1994HIRES:Telescope}. HARPS (S1D) and UVES one-dimensional merged spectra were downloaded from the ESO archive. We also obtained new Keck/HIRES spectra of M dwarfs with available UV spectra but no Ca II H\&K spectra. We were awarded 3 half-nights (2019-03-01, 2019-07-07, and 2019-07-08). We used the HIRES blue arm with the C5 decker (1.148\arcsec~$\times$~7\arcsec~slit), the KV370 filter, echelle angle = 0.046918$^{\circ}$, cross-dispersion angle = 1.9523$^{\circ}$, and standard 2$\times$1 binning. The nominal spectral resolving power of this mode is R=37,000. We obtained new spectra covering wavelengths 3750-6720 \AA~for 4 M dwarfs (LP 5-282, G 75-55, LP 247-13, GJ 3290), and we use the data products from the automatic pipeline \texttt{MAKEE}\footnote{http://www.astro.caltech.edu/~tb/makee/} as was used for all archival HIRES spectra used in this work. 
For targets with no available optical spectra with sufficient signal-to-noise from HARPS, HIRES, or UVES, we used single spectra available to our team from the Canada-France-Hawaii Telescope ESPaDOnS spectrograph\footnote{http://www.cadc-ccda.hia-iha.nrc-cnrc.gc.ca/en/cfht/} (2MASSJ02001277-0840516 and GJ3997) and the Observatoire de Haute-Provence ELODIE spectrograph\footnote{http://atlas.obs-hp.fr/elodie/} (2MASSJ04223953+1816097 and 2MASSJ04184702+1321585; \citealt{Moultaka2004The1}). 

\begin{figure}
\centering
\includegraphics[width=12cm]{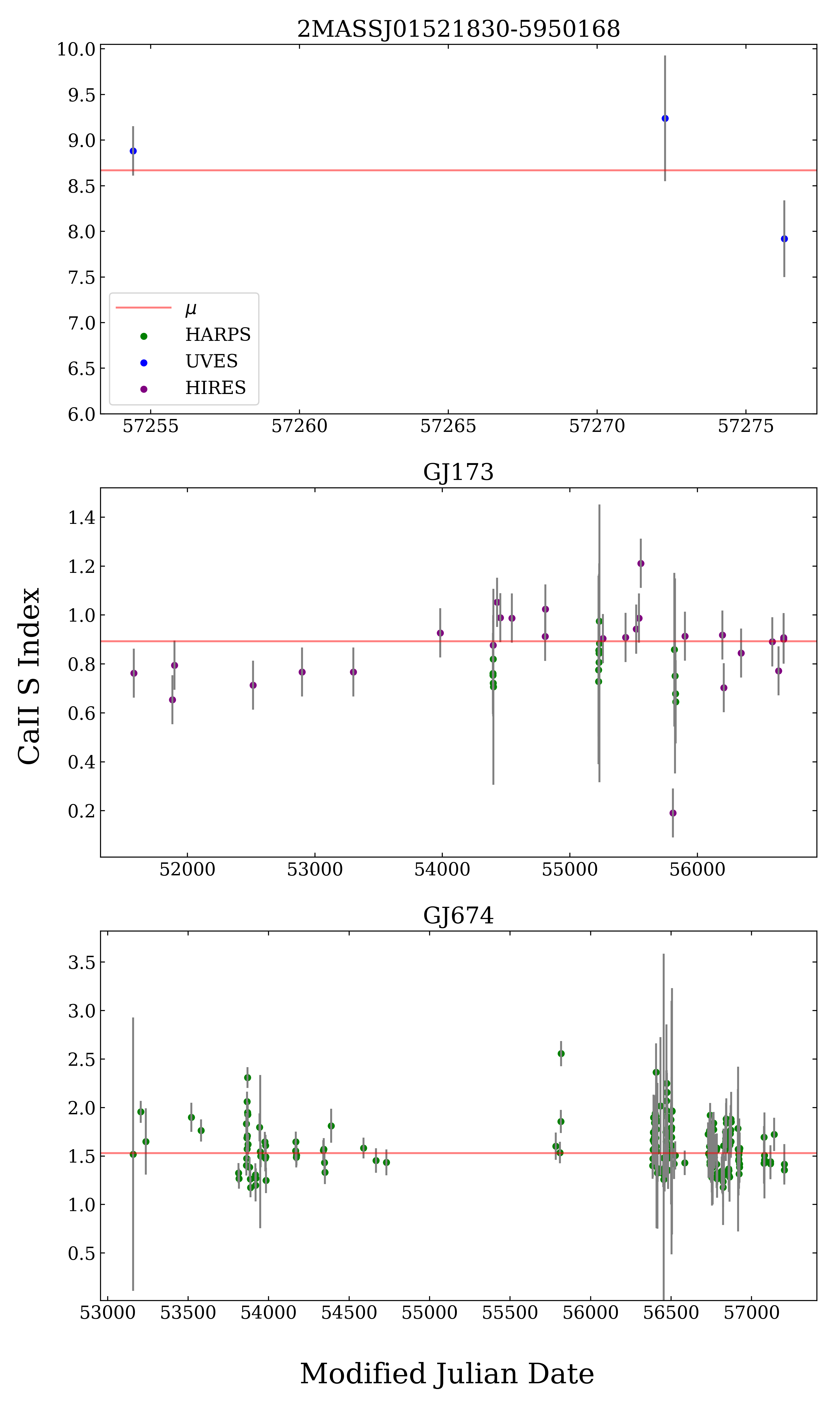}
\caption{Stars representing the range of quantity and quality of S index measurements available for the target stars in our sample. Some, including 2MASSJ01521830-5950168 (top panel), have only a few observations in total. Few have been observed by multiple instruments, but GJ173 is an example of a star observed with both HIRES and HARPS. Finally, GJ674 shows how some stars have multiple observations in a short amount of time with longer spans without any observations. Many of the values have error bars smaller than the points.}
\label{time_var_samples}
\end{figure}

We did not flux calibrate any of our optical spectra; the activity indicators used rely on normalizations to nearby continuua, and we compared our measurements to overlapping samples from \citealt{Newton2017TheRotation} and \citealt{Astudillo-Defru2016MagneticRHK} to ensure our technique is in line with theirs. To account for the time variability of stellar activity levels with non-contemporaneous observations, we found the optical activity indices of each spectrum individually and took the signal-to-noise weighted average for the final result (described in Sections~\ref{sec:HaEquivalentWidths} and \ref{sec:CaIIIndices}). Errors calculated for the optical parameters of our target sample do not take into account stellar variability, and not all stars were observed over multiple epochs. Measurements for a representative sample from our target list are shown in Figure \ref{time_var_samples}. 

\subsection{Ultraviolet Data}\label{subsec:UV}

Ultraviolet spectra came from the Hubble Space Telescope (HST) Space Telescope Imaging Spectrograph (STIS) and Cosmic Origins Spectrograph (COS). A variety of gratings and central wavelength settings (noted as grating with central wavelength setting in parentheses) were used in this diverse data set, including G140L (1425 \AA{}), G140M (1222 \AA{}), E140M (1425 \AA{}), E140H (1271 \AA{}), G230L (2950 \AA{}), E230M (2707 \AA{}), and E230H (2713 \AA{}) for STIS, and G140L (1105, 1230, and 1280 \AA{}), G130M (1222, 1291, 1309, 1318, and 1327 \AA{}), G160M (1533, 1577, 1589, 1600, 1611, and 1623 \AA{}), and G230L (2376, 2635, 2950, and 3360 \AA{}) for COS. We used spectra downloaded from MAST reduced with the standard STScI pipeline, except for spectra from the MUSCLES and Mega-MUSCLES Treasury Surveys (Wilson et al. 2020 In Review, \citealt{France2016TheOverview,Youngblood2016TheExoplanets,Loyd2016ThePlanets,2019ApJ...871L..26F}), FUMES (Pineda et al. Under Review), and the STARCat catalog \citep{Ayres2010StarCAT:STARS}; see references for descriptions of the data reduction. Similarly, data from GO 15190 and GO 15326 were reduced using the same methods used in the STARCat catalog.  

For the UV spectra, we took the error-weighted average of all available spectra before measuring each spectral line, because the signal-to-noise of each UV spectrum was generally lower. To measure the flux of the UV emission lines not found in previous literature (noted in Table~\ref{table:rhk_correlations}), we fit a Voigt profile, because it accurately captures the core and the wings of each line and requires only four parameters, allowing for computational efficiency. For the pair of blended lines that comprise Si II and Mg II, we fit a double Gaussian because the wings of these lines are typically below the instrumental noise floor. Uncertainty for each Voigt parameter and the integrated line flux were determined by bootstrapping: we randomly sampled with replacement the flux values for each line and fit a Voigt profile to each new sample. Although we considered block bootstrapping -- resampling data pairs with replacement in subsections of each spectral line range that have similar noise profiles -- we determined there was no effect on our results and therefore resampled across the entire wavelength range for each spectral line. We report the median and standard deviation of the set of Voigt integrals flux and the error. In Section~\ref{sec:UVOpticalRelations}, we use the UV line luminosities normalized to the stellar bolometric luminosity, which we calculate as L$_{\text{bol}}$ = $\sigma_{SB}T_{\text{eff}}^4~4\pi R_{\star}^2$ using stellar parameters from Table \ref{table:machinereadabletable}. Uncertainties in our $T_{eff}$ and $R_{\star}$ parameters range from 1-5\% and 1-10\%, respectively. In particular, the largest stars in our sample (2MASSJ01521830-5950168, 2MASSJ03315564-4359135, and 2MASSJ23261069-7323498) have $1.07\pm-0.17$ R$_{\odot}$ \citep{Malo2014BANYAN.MODELS}.

Absorption from gaseous species in the interstellar medium has at least some effect on the majority of the UV emission lines analyzed, including \Lya, \MgII, \CII, \SiII, and \SiIII. For Si III, the effect is expected to be negligible \citep{Redfield2004TheParsecs}, so corrections were not applied. The impact on the Si II emission line at 1260 \AA~by the ISM ranges from negligible to moderate depending on the line of sight  \citep{Redfield2004TheParsecs}. No correction was applied, and any uncorrected ISM absorption may cause some scatter in our Si II correlations. ISM effects are most noticeable for the \Lya{} line with complete absorption in the line center caused by optically thick \HI. We only use intrinsic \Lya{} fluxes that have been reconstructed from their observed profiles \citep{Wood2005StellarAstrospheres, Youngblood2016TheExoplanets, Youngblood2017TheDwarfs} with typical uncertainties of 5-30\%. The \CII{} 1334 \AA{} line is similarly affected by ISM absorption, so only the flux from the 1335 \AA{} line of the CII doublet is included in the listed \CII{} flux values. Additionally, both lines in the Mg II doublet experience attenuation, which is accounted for by assuming a uniform $30\pm10$\% correction to measured values (propagating errors accordingly) based on assumptions discussed by  \cite{Redfield2002TheParsecs} and \cite{Youngblood2017TheDwarfs}. With regard to our optical lines, CaII can also be absorbed by Ca$^+$ in the ISM, but attenuation is only significant for stars at a distance beyond 100 pc \citep{Fossati2017TheStudies}, of which there are none in this study.

\subsection{H$\alpha$ Equivalent Widths and \Lhabol}\label{sec:HaEquivalentWidths}

We measured \Ha{} equivalent widths (EW) using the equation
\begin{equation} \label{ew_eqn}
\text{EW} = \int_{\lambda_1}^{\lambda_2}\Big(1-\frac{F_\lambda}{F_c}\Big)d\lambda,
\end{equation}

\noindent where $F_\lambda$ is the flux of each wavelength across the width of the line and $F_c$ is defined as the average continuum from two ranges on either side of the line,  $6500-6550$\AA{} and $6575-6625$\AA{}. For $d\lambda$, we use the average pixel width between each observed wavelength in a spectrum, as variations are negligible. We follow the method used by \cite{West2011THEDATA} and \cite{Newton2016TheRotation} and assign the bounds of integration to be $\lambda_1=6558.8$\AA{} and $\lambda_2=6566.8$\AA{} for all spectra. The equivalent width weighted means and uncertainties for all targets are listed in Table \ref{table:machinereadabletable}.

We validate our EW measurements by comparing with \cite{Newton2016TheRotation} \Ha{} EWs ($H\alpha_{N}$) for 12 overlapping targets, finding the best fit line $H\alpha_{N} = 1.05(\pm0.003)H\alpha + 0.04(\pm0.03)$. The slight departure from a precise 1:1 line is a result of the use of multiple instruments in our calculations, and no additional calibration was performed to match our measured values to Newton et al. When optical spectra were not available for \Ha{} measurements, we used literature values from \cite{Riaz2006IdentificationNeighborhood} and \cite{Malo2014Banyan.Groups}, that used the IRAF splot package to calculate \Ha{} equivalent widths for a sample of M dwarfs. Our calculated values are comparable to those found in literature ($H\alpha_{lit}$). Fitting a least-squares linear regression, we found a best fit line of $H\alpha_{lit} = 0.98(\pm0.04)H\alpha - 0.34(\pm0.41)$ and therefore we assume no significant variation in \Ha{} values across both studies and this work.

In order to remove the stellar mass dependence of our H$\alpha$~index (the Wilson-Bappu effect; \citealt{Wilson1957HTOPICS,Stauffer1986CHROMOSPHERIC1}), we calculate \Lhabol{} using the methods of \cite{2014ApJ...795..161D} and \cite{Newton2017TheRotation}. First we subtract from our EW measurements the the minimum $H\alpha$ EW value for a star of a given mass using relations from \cite{Newton2017TheRotation}. We estimate the masses of our targets to accuracies of 10-20\% from various literature sources. For the few cases where literature masses were unavailable, we used BT-Settl isochrones\footnote{https://phoenix.ens-lyon.fr/Grids/BT-Settl/CIFIST2011\_2015/ISOCHRONES/} \citep{Allard2011ModelsExoplanets} with solar abundances \citep{Caffau2010SolarAtmosphere} and the target's effective temperature (Table~\ref{table:machinereadabletable}) to determine a mass. We then converted these corrected \Ha{} EWs into L$_{\text{H}\alpha}$/L$_{\text{Bol}}$ by multiplying each EW by a spectral type conversion factor ($\chi$) determined by \cite{2014ApJ...795..161D}. 

\subsection{\cahk{} indices}\label{sec:CaIIIndices}

\subsubsection{The S index}\label{subsubsec:Sindex}

\begin{figure}
\centering
\includegraphics[width=18cm]{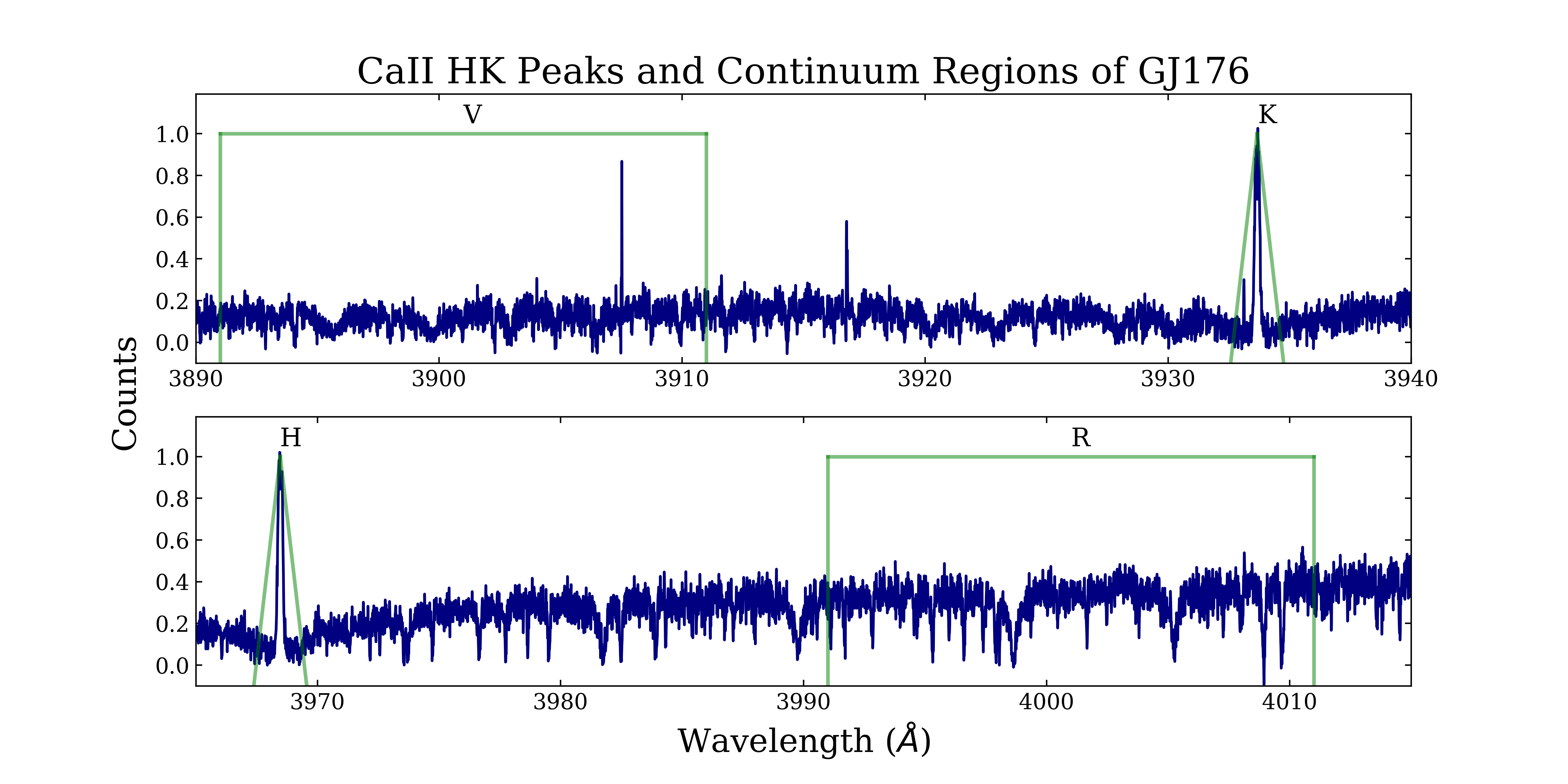}
\caption{Demonstration of continua and triangular passbands used in the \ca{} analysis \citep{Wilson1968FLUXK-LINES, Vaughan1978FluxEmission}. The Ca II H line (3968.47 \AA) passband and continuum regions avoid the H$\epsilon$ emission line (3970.07 \AA). This figure uses an individual S1D spectrum of GJ 176 observed with HARPS as an example.}
\label{CaHK_example}
\end{figure}

The S index measures the flux ratio between the \cahk{} lines and the surrounding continuum, standardized by an instrumental calibration factor \citep{Vaughan1978FluxEmission}. Two triangular bandpasses are centered on the H and K lines with 1.09 \AA{} FWHM and two 20 \AA{} top-hat bandpasses are centered at 3901 \AA{} and 4001 \AA{}, known as the V and R ranges, respectively (Figure \ref{CaHK_example}). The S index is limited in comparing activity levels across spectral types; as the integrated continuum emission decreases for cooler stars, the S index will increase \citep{Middlekoop1982MagneticStars}. In addition, the triangular bandpasses include contaminating photospheric emission in addition to the desired chromospheric emission.

We follow the method in \cite{Lovis2011TheVelocities} and \cite{Astudillo-Defru2016MagneticRHK} to allow for optimal validation of our measurements, as there is significant overlap of our dataset with theirs. Our measured values are compiled in Table \ref{table:machinereadabletable}. We calculate the S index for all of our target stars by determining the mean flux emitted across both 20 \AA{} continuum regions and the 1.09 \AA{} regions centered on each line instead of integrating over them.

\begin{equation} \label{our_s_index}
S = \alpha \frac{\tilde{f_H} + \tilde{f_K}}{\tilde{f_V} + \tilde{f_R}},
\end{equation}

\noindent where $\alpha \approx 1$ and $\tilde{f_V}$, $\tilde{f_R}$, $\tilde{f_H}$, and $\tilde{f_K}$ are the mean flux emitted in the 20 \AA{} continuum regions (V,R) and the 1.09 \AA{} line regions (H,K). To confirm the accuracy of our measurements, we fit a linear least-squares regression of our S index values to the Mount Wilson values provided by \citealt{Astudillo-Defru2016MagneticRHK} (22 overlapping targets total) to find the following relation: $S = 1.05(\pm0.01) S_{orig} + 0.06(\pm0.01)$, with $S$ representing the S indexes found in their study and $S_{orig}$ denoting those found in our work. For this comparison, we used S index values found from all three ground-based spectrographs: HARPS, UVES, and HIRES. There were not enough overlapping observations obtained with these three spectrographs to quantify the differences between the measured S index of individual targets using multiple instruments.

\subsubsection{\rhk{}}\label{subsubsec:rhk}

\newpara

Accurate characterization of the chromospheric activity levels is essential for this work. UV radiation originates in magnetically-heated regions of the stellar atmosphere above the photosphere, meaning photospheric emission is not correlated with UV emission lines. Thus we measure the \rhk{} index \citep{Middlekoop1982MagneticStars, Rutten1984MagneticStars, Noyes1984RotationStars}, which includes corrections for the photospheric flux in the continuum (V,R) and line (H,K) regions of the S index. $R'_{HK}$ has been well-defined for F, G, and K spectral types \citep{Lovis2011TheVelocities} and has been previously extrapolated to later spectral types with less accuracy (as discussed in \citealt{Astudillo-Defru2016MagneticRHK}). Past attempts to characterize $R'_{HK}$ for M dwarfs used the original $B-V$ color index \citep{Mittag2013AstronomyAtmospheres}, which is not best suited for M dwarfs due to their frequent lack of available B band photometry in the literature and V band sensitivity to metallicity \citep{Astudillo-Defru2016MagneticRHK, Delfosse1998RotationDwarfs, Bonfils2013AstrophysicsSample}. 

\rhk{} is related to the S index by

\begin{equation} \label{r_prime_hk}
R'_{HK} = R_{HK} - R_{phot} = K \sigma_{\text{SB}}^{-1} 10^{-14} \text{C}_{cf} (S-S_{phot}),
\end{equation}

\noindent where $R_{phot}$ and $S_{phot}$ are photospheric contributions to $R'_{HK}$ and $S$ respectively, $\sigma_{\text{SB}}$ is the Stefan-Boltzmann constant, C$_{cf}$ is the color correction factor, and $K$ is a factor that transforms arbitrary fluxes to surface fluxes. \cite{Middlekoop1982MagneticStars} and \cite{Rutten1984MagneticStars} both calculated $K$, but $1.07 \times 10^6$ erg cm$^{-2}$ s$^{-1}$ is the most recent value provided in \citealt{Hall2007TheObservations}. 

In calculating the color correction factor, C$_{cf}$, and $R_{phot}$, we follow Equation 9, Equation 10, and Table 1 from \cite{Astudillo-Defru2016MagneticRHK}. We elect to use the coefficients provided corresponding to $V-K$ color index in the Johnson photometic system \citep{Johnson1966AstronomicalInfrared}, as V band fluxes are generally more available than I band fluxes. The relations between $V-K$, C$_{cf}$, and R$_{phot}$ are: 

\begin{equation} \label{log_ccf}
\log \text{C}_{cf} = -0.005(V-K)^3 + 0.071(V-K)^2 -0.713(V-K) + 0.973
\end{equation}
\begin{equation} \label{log_rphot}
\log \text{R}_{phot} = -0.003(V-K)^3 + 0.069(V-K)^2 - 0.717(V-K) -3.498.
\end{equation}

A small number of our stars have no V band magnitudes available in the literature, and we calculate the V magnitude from V-K estimates provided by \cite{Pecaut2013INTRINSICSTARS}. Uncertainties are assumed to be 5\% of the flux value; we assumed the same relative error for stars whose V magnitudes have no published uncertainties.

\section{UV-optical Relations}\label{sec:UVOpticalRelations}
\begin{figure}
\centering
\includegraphics[width=12cm]{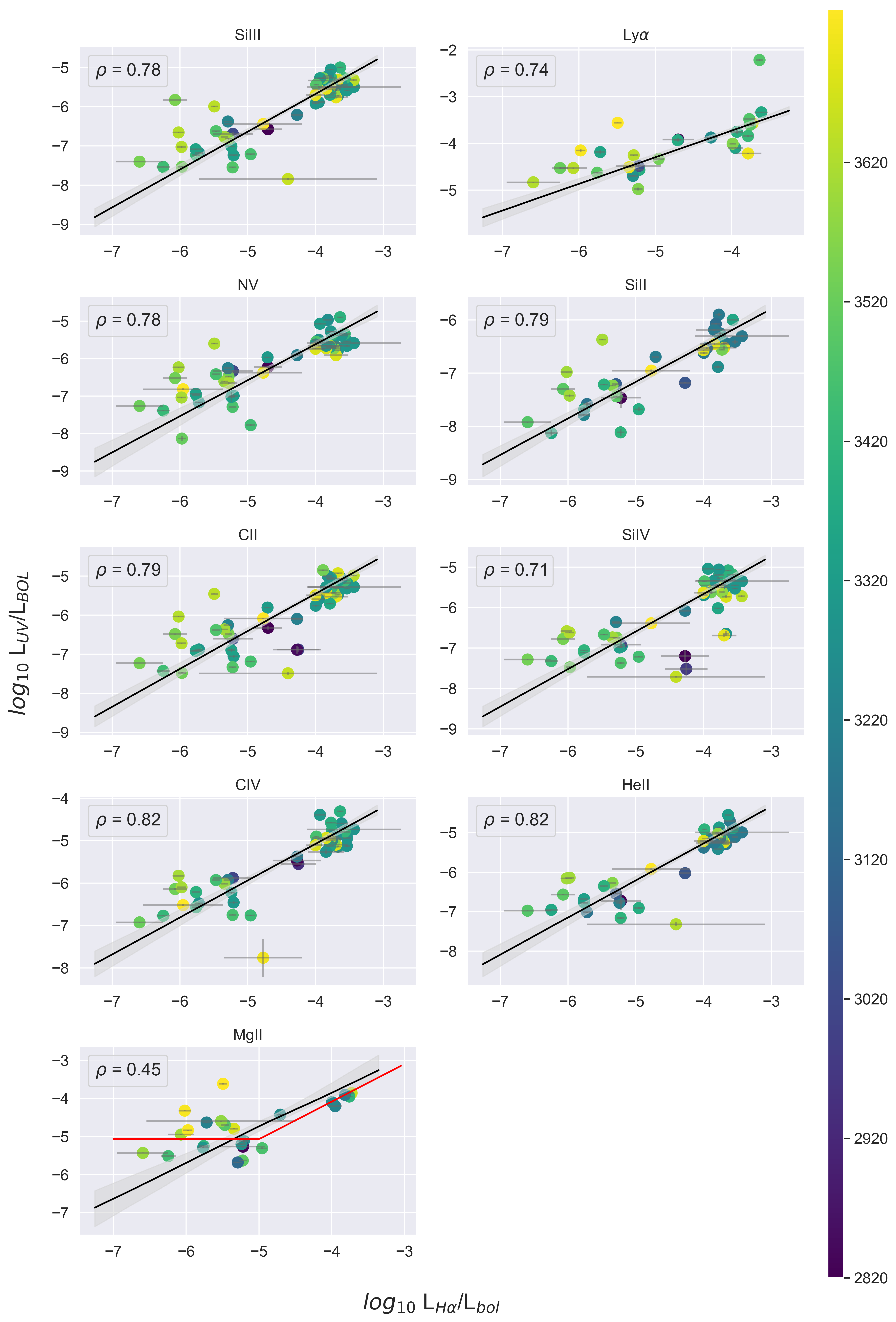}
\caption{logL$_{\text{UV}}$/logL$_{\text{Bol}}$ as a function of \Lhabol{}. Each circle represents a different stellar target from our sample color-coded by effective temperature. The logL$_{\text{UV}}$/logL$_{\text{Bol}}$ label on the Y-axis refers to the normalized UV luminosity of each individual emission line, and the luminosity errors are present but do not extend past the edges of each point. Calculated Spearman rank correlation coefficients ($\rho$) are shown for each graph. The black lines show the best-fits (see Table \ref{table:ha_correlations} for parameters), and the grey shaded regions show the 1$\sigma$ errors on the fits. In the \Lya~subplot, the green point (LP 247-13) at \Lhabol{} = -3.64$\pm$0.05, log$_{10}$ L(\Lya)/L$_{\rm Bol}$ = -2.22$\pm$0.02 is excluded from the fit. In the Mg II subplot, the yellow point (GJ 676 A) at \Lhabol{} = -5.49$\pm$0.05, log$_{10}$ L(MgII)/L$_{\rm Bol}$ = -3.62$\pm$0.02 has been excluded from the black line fit, and we present an additional broken power law fit (red line) that may better fit the data. Both fits are quantified and described further in Table \ref{table:ha_correlations}. A version of this figure without the best fit lines and with the individual star names labeled is available in the online journal.}
\label{Ha_vs_lum}
\end{figure}

\begin{figure}
\centering
\includegraphics[width=14cm]{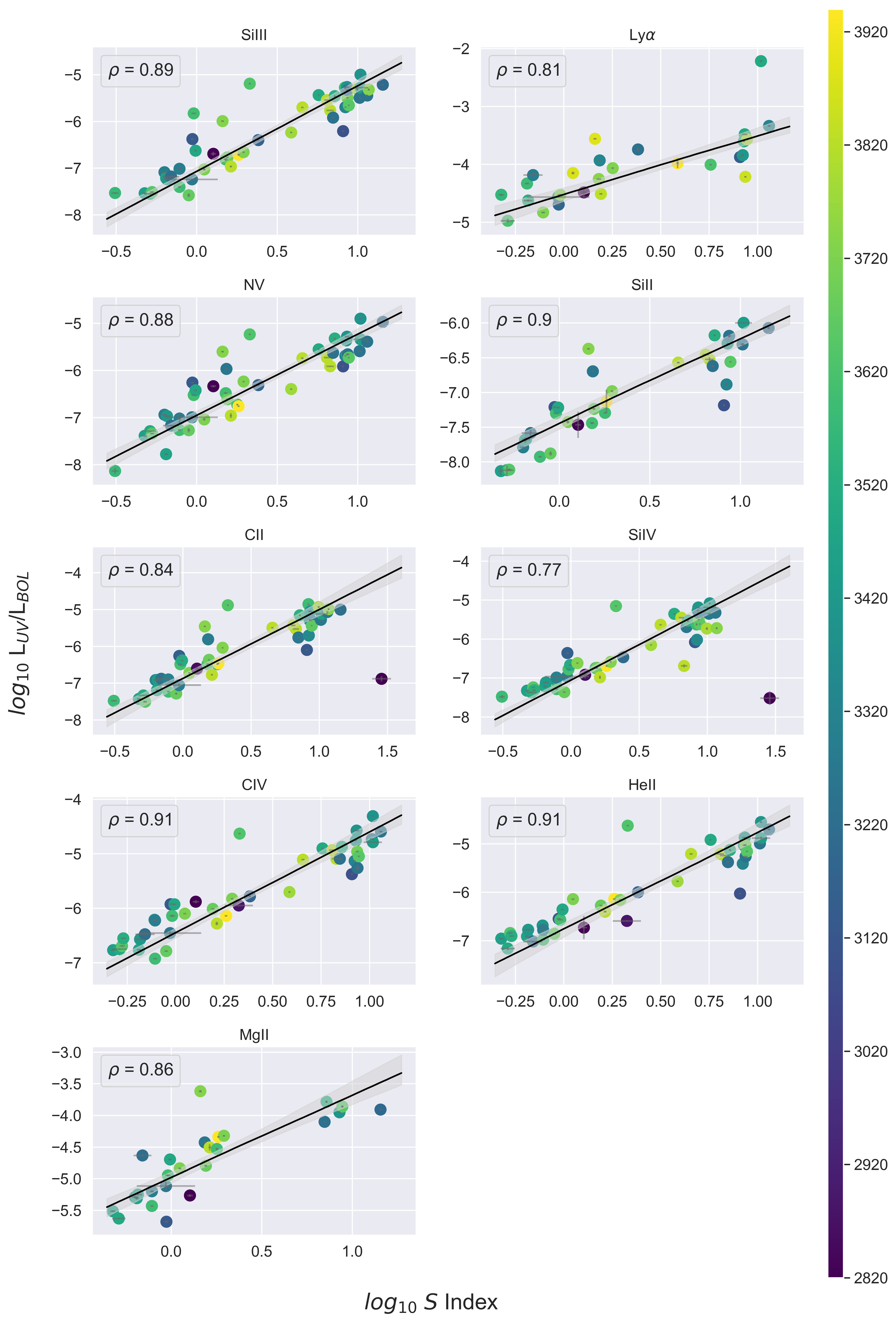}
\caption{Same as Figure~\ref{Ha_vs_lum} for logL$_{\text{UV}}$/logL$_{\text{Bol}}$ as a function of the Ca II S index. A version of this figure without the best fit lines and with the individual star names labeled is available in the online journal.}
\label{s_vs_lum}
\end{figure}

\begin{figure}
\centering
\includegraphics[width=13cm]{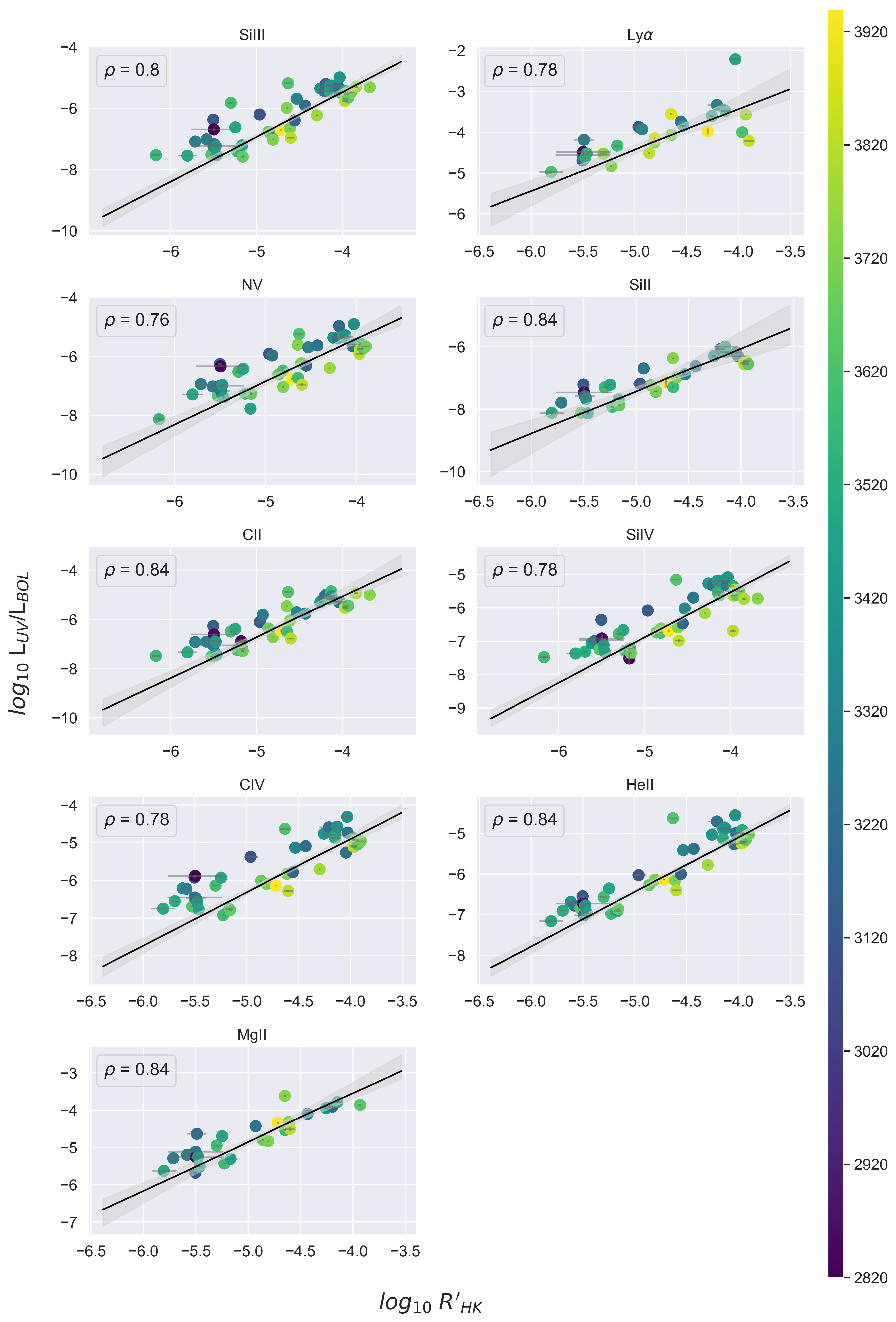}
\caption{Same as Figure~\ref{Ha_vs_lum} for logL$_{\text{UV}}$/logL$_{\text{Bol}}$ as a function of the Ca II \rhk{} index. A version of this figure without the best fit lines and with the individual star names labeled is available in the online journal.}
\label{rhk_vs_lum}
\end{figure}

We analyzed the relation between each optical activity indicator and the luminosities of nine different UV lines each normalized by the bolometric luminosity (listed in Table \ref{table:machinereadabletable}). For \Lhabol, S index, and \rhk, power laws were fitted to the data (including uncertainties) in log space using the same bootstrapping method described in Section~\ref{sec:ObservationsReductions}. We re-sampled the data with replacement and found a best-fit line each time, then found the median and the 68\% confidence interval among all the fits performed, shown in Figures \ref{Ha_vs_lum}, \ref{s_vs_lum}, and \ref{rhk_vs_lum}. Tables \ref{table:rhk_correlations}, \ref{table:ha_correlations}, and \ref{table:s_correlations} list the fitted power law parameters for the median line and 1-$\sigma$~uncertainties as well as the Spearman rank-order correlation coefficients ($\rho$), and the standard deviations about the best fit line.

Figure \ref{Ha_vs_lum} shows the relation between \Lhabol{} and logL$_{\text{UV}}$/logL$_{\text{Bol}}$ for each line, and Table \ref{table:ha_correlations} describes the fit parameters. We find statistically significant, positive correlations between \Lhabol{} and logL$_{\text{UV}}$/logL$_{\text{Bol}}$ for all UV emission lines. There are two regions of stars, active (H$\alpha$ in emission) and inactive (H$\alpha$ in absorption). The active stars are clustered in the area of each graph in Figure \ref{rhk_vs_lum} with \Lhabol{} $> \sim-4$, and the inactive stars are scattered along the rest of the best fit lines. When analyzed separately, there was no significant correlation in either region; however, our sample spans a wide range of H$\alpha$ values, which demonstrate a significant correlation when examined all together. Mg II has the weakest correlation with \Lhabol{} and most uncertain best-fit line of all the UV emission lines, even when excluding the outlier GJ 676 A (Section~\ref{subsec:outliers}) from the fit, due to a lack of stars in our sample that are active in both Mg II and H$\alpha$. This is likely because of the recent HST brightness restrictions for M dwarfs that are most strict in the NUV regime, affecting the community's ability to collect Mg II data for more active stars. We present in the caption of Table~\ref{table:ha_correlations} an alternate fit that parameterizes the apparent flattening of L(MgII)/L$_{bol}$ in the inactive regime. We have also removed the outlier LP 247-13 from the L(\Lya)/L$_{bol}$--\Lhabol{} fit because it drives the best-fit slope to a steep value that does not match the other active stars. 

Here we describe the comparison between \Ha{} EW and logL$_{\text{UV}}$/logL$_{\text{Bol}}$, which is not shown because there is no apparent correlation. The targets are divided into two loci: an active and inactive regime. As the equivalent width becomes more negative (more active), UV luminosity plateaus at a nearly constant high value. For smaller equivalent widths (EW $\sim$ 0; less active), there is a large spread of several orders of magnitude in normalized UV luminosity. Our sample has 52 M dwarfs from M0-M5.5 with \Ha{} equivalent widths concentrated near the threshold between active ($<-1$\AA{}) and inactive ($>-1$\AA{}) M dwarfs. \Ha{} absorption lines deepen with increased activity before flipping to emission \citep{Cram1985FORMATIONSTARS}, leading to a non-monotonic relation between stellar activity and \Ha{} equivalent width whereas UV emission scales monotonically with stellar activity. This mostly affects stars with \Ha{} equivalent widths $\gtrsim -1$ \AA{}, making \Ha{} EW a poor activity indicator for inactive M dwarfs \citep{Walkowicz2008TracersDwarfs}.

The relationships between the logS index and logL$_{\text{UV}}$/logL$_{\text{Bol}}$ line emissions are shown in Figure \ref{s_vs_lum}. The S index includes photospheric contamination that leads to a wider spread of UV flux values at each S index, as the UV lines do not originate in the photosphere (e.g., \citealt{1981ApJS...45..635V}). Because of this, we did not expect a tight correlation with UV luminosity, but the Spearman rank-order correlation coefficients indicate a statistically significant positive correlation for each emission line. Parameters for the best fit lines are shown in Table \ref{table:s_correlations}. The scatter about the S index best fit lines are greater than about the \Lhabol{} best fit lines for all emission lines except Mg II.

Transforming the S index to \rhk{} removes the unwanted photospheric contribution that is present in the S index. Because the UV emission lines studied originate in regions of the stellar atmosphere that are above the photosphere and dominated by magnetic heating, we expected to find correlations with less scatter between log\rhk{} and logL$_{\text{UV}}$/logL$_{\text{Bol}}$ for each emission line. Figure \ref{rhk_vs_lum} shows statistically-significant correlations between log\rhk{} and each of the normalized UV emission line luminosities. The scatter about the log\rhk{} best fit lines is significantly smaller than for both \Lhabol{} and S index, except for N V, where the \Lhabol{} has a smaller scatter by 0.07 dex.

For each optical activity index, all of the individual UV lines' power law slopes are consistent with each other within 1-$\sigma$ uncertainties. For \rhk, S index, and \Lhabol, the weighted averages are, respectively, 1.57$\pm$0.06, 1.89$\pm$0.07, and 0.96$\pm$0.15. The Mg II fit dominates the uncertainty in the \Lhabol~average power law slope. Excluding Mg II, the weighted average power law slope for \Lhabol{} becomes 0.95$\pm$0.04. We find no correlation between power law slope and line formation temperature for any of the fits. The Spearman correlation coefficient indicates a strong positive correlation with values for each UV emission line ranging between 0.76 $\leq \rho \leq$ 0.85. The probability of no correlation ($n$) is $<10^{-6}$ for each fit. The standard deviations about the best fit lines are 0.31 $\leq \sigma \leq$ 0.61 dex.  Comparing these parameters to \citealt{Youngblood2017TheDwarfs}, we see that our standard deviations about the best fit are generally larger, indicating more scatter. This is likely due to our larger and more diverse sample of stars in addition to using a different Ca II activity indicator (log\rhk{}). Additionally, we analyzed the relationship between the residuals around our log\rhk{} best fit for each UV line and stellar effective temperature and found a shallow, statistically-significant negative correlation. This indicates that scatter in the correlation increases slightly for higher stellar temperatures.

\subsection{Outliers} \label{subsec:outliers}

Despite a tight relation between \rhk{} and logL$_{\text{UV}}$/logL$_{\text{Bol}}$ across 3 orders of magnitude of logL$_{\text{UV}}$/logL$_{\text{Bol}}$ values in our sample of M dwarfs, some stars are not well fit by the established trends. These include GJ849, GJ3290, GJ876, LP247-13, 2MASSJ23261069-7323498, and GJ 676 A. Almost all of these stars are outliers in the sense that their normalized UV line luminosity is much higher than other stars of a similar \rhk{} value, except for 2MASS J23261069-7323498. We have examined all of these stars' spectra for flares and found none that could explain such a large discrepancy from their neighbors. GJ 876 (M4V; P$_{rot}$ = 96.7 days; \citealt{Rivera2005}) was noted as a UV-bright outlier in \cite{Youngblood2017TheDwarfs}, and had many strong flares \citep{Youngblood2017TheDwarfs,Loyd2018MUSCLESPaperV}. While we do not think that large flares are affecting the UV line measurements, they are indicative of an elevated activity that may manifest itself in other ways, such as the quiescent UV luminosity. LP 247-13 is a young M3 dwarf and a potential member of the $\sim$625 Myr old Hyades cluster, although \cite{Shkolnik2012IDENTIFYING} note that its low surface gravity indicates it could be much younger. This star is one of the most UV-bright members of our sample, and in particular, its L(Ly$\alpha$)/L$_{bol}$ is an order of magnitude greater than the next brightest star. LP 247-13 significantly drives the slope of the \Lya--\rhk{} correlation to a steeper value, and was removed from the \Lya--\Ha{} fit for its extreme effect on that best fit line. More details about LP 247-13 can be found in upcoming publications about the FUMES survey (Pineda et al., in preparation and Youngblood et al., in preparation). Similarly, GJ 3290 and 2MASSJ23261069-7323498 are both young stars (625 Myr and 40 Myr, approximately) and will be discussed in an upcoming publication about the HAZMAT survey (Loyd et al., in preparation). GJ 849 and GJ 676 A are both field age stars and will be discussed in upcoming publications about the Mega-MUSCLES survey. 

Here we discuss the impact of stellar variability on the general scatter in our correlations. None of our targets' optical and UV spectra were taken simultaneously, and the time between the spectra span days to years. M dwarfs exhibit activity-related variability in UV and optical emission lines on many timescales including minutes (flares; \citealt{Baliunas1984,Hilton2010,Loyd2018MUSCLESPaperV}), days-to-months (rotational modulation and the emergence and decay of active regions and starspots; \citealt{Vaughan1981,SuarezMascareno2015}), and years (magnetic activity cycles; \citealt{GomesdaSilva2012,Robertson2013,Buccino2014,Toledo2019}). Mitigating the effect of rotation and stellar cycles through modeling a base line for the sample is a promising topic for future work. It would also be interesting but extremely challenging to gather a truly simultaneous UV + optical dataset to remove the scattering effect of stellar variability within individual stars from UV-optical correlations.

We have removed the effects of large flares by visually inspecting our optical spectra and removing spectra exhibiting obvious flaring (significantly brighter emission lines and/or continuum enhancement). We inspected the UV light curves of outliers for flares, and we did not find any that could have increased the UV line luminosity by a factor of 2 (0.3 dex) or more. Typically, integration times in the UV are $\gtrsim$1 hour long to build up sufficient S/N, and brief flares do not greatly affect the cumulative spectrum \citep{Loyd2018MUSCLESPaperV,Loyd2018HAZMAT.Ultraviolet}. Our significant outliers (listed above), have UV line luminosities that are $>$1 dex above or below the best-fit line, which is not explained by flares. However, at least some of the scatter in our correlations must be due to flares as we were unable to discern small flares in our spectra.

Almost all of our optical and UV data were taken non-simultaneously, which means that our results are susceptible to rotational and activity cycle effects that occur on $>$1 day timescales. This affects our UV spectra the most as almost all of our targets had their UV data taken in a single day, whereas many of our targets' optical spectra were taken over multiple epochs spread over years and decades. Thus, the optical spectral variability over rotational and activity timescales should be averaged out. Assuming the rotational and cycle variability of M dwarfs is similar to the Sun's, the amplitude of variations are significantly larger in the UV than they are in the optical. Based on SORCE solar spectral irradiance timeseries data\footnote{https://lasp.colorado.edu/lisird/} \citep{McClintock2005}, the Sun varies on 27-day (rotational) timescales by 1-5\% at solar minimum and 10-30\% at solar maximum in UV lines like C II, N V, and \Lya, although \Lya~can vary by as little as 5\% during solar maximum. Over the course of the 11-year solar cycle, the same lines modulate by $\sim$30\%, although \Lya~can modulate by as much as 50\%. Conversely, the Sun's S-index and \rhk{} values vary on an 11-year timescale from log S = -0.80 to -0.74 and log \rhk{} = -4.98 to -4.91, or $\sim$15\% (0.06-0.07 dex) \citep{Egeland2017}. This variation is much smaller than the error bars on our measurements. M dwarf activity cycles have been detected in the UV for GJ 551 (Proxima Centauri); \citep{Wargelin2017} found $\sim$10\% amplitude variations in the broadband Neil Gehrels Swift Observatory's UVW1 (NUV) filter over several years. Tracing M dwarf cycles in the optical is more common, with 2-3\% fluctuations observed in H$\alpha$~intensity \citep{Robertson2013} and 10-30\% fluctuations observed in Ca II H\&K intensity \citep{Buccino2014,Toledo2019}. 

\begin{deluxetable}{cccccccc}
\tabletypesize{\small}
\tablecolumns{8}
\tablewidth{0pt}
\tablecaption{ Fit parameters for log$_{10}$ \rhk{} and log$_{10}$ L$_{UV}$/L$_{BOL}$ \label{table:rhk_correlations}} 
\tablehead{\colhead{Transition name} & 
                  \colhead{Wavelength (\AA)} & 
                  \colhead{log $T_{\rm formation}^a$} & 
                  \colhead{$\alpha$} &
                  \colhead{$\beta$} &
                  \colhead{$\rho$} & 
                  \colhead{$n$} &
                  \colhead{$\sigma$}
                  }
\startdata
SiIII & 1206.50 & 4.7 & 1.47$\pm$0.13 & 0.42$\pm$0.61 & 0.80 & $<$0.0001 & 0.50 \\
LyA & 1215.67 & $\lesssim$4.5 & 1.07$\pm$0.19 & 0.92$\pm$0.89 & 0.78 & $<$0.0001 & 0.44 \\
NV & 1238.82, 1242.8060 & 5.2 & 1.53$\pm$0.21 & 0.8$\pm$0.99 & 0.76 & $<$0.0001 & 0.58 \\
SiII* & 1260.42, 1264.74, 1265.00 & 4.5 & 1.45$\pm$0.26 & -0.16$\pm$1.20 & 0.84 & $<$0.0001 & 0.47 \\
CII$^b$ & 1335.71 & 4.5 & 1.73$\pm$0.20 & 1.96$\pm$0.96 & 0.84 & $<$0.0001 & 0.56 \\
SiIV & 1393.76, 1402.77 & 4.9 & 1.36$\pm$0.11 & -0.07$\pm$0.52 & 0.78 & $<$0.0001 & 0.49 \\
CIV & 1548.19, 1550.78 & 5.0 & 1.43$\pm$0.13 & 0.86$\pm$0.63 & 0.78 & $<$0.0001 & 0.48 \\
HeII & 1640.4$^c$ & 4.9 & 1.35$\pm$0.09 & 0.32$\pm$0.41 & 0.84 & $<$0.0001 & 0.35 \\
MgII*$^{,d}$ & 2796.35, 2803.53 & $\lesssim$4.5 & 1.38$\pm$0.17 & 2.03$\pm$0.80 & 0.84 & $<$0.0001 & 0.36 \\
\enddata
\tablecomments{The scaling relations take the form log$_{10}$ $L_{\rm UV}$ = ($\alpha$ $\times$ log$_{10}$ \rhk{}) + $\beta$, where $L_{\rm UV}$ represents each UV emission line luminosity in erg s$^{-1}$. $\rho$~is the Spearman correlation coefficient, $n$ is the probability of no correlation, and $\sigma$~is the standard deviation of the data points about the best-fit line (dex).}
\tablenotetext{*}{Fit with a double Gaussian because wings cannot be resolved at the lower signal to noise of these blended emission lines.}
\tablenotetext{a}{Formation temperatures are from the CHIANTI database \citep{Dere1997CHIANTILines,Landi2013CHIANTIANCHANGES}. Note that the Ca II H\&K line cores form around 10$^{3.8}$ K, a similar temperature to the \Lya~and Mg II~line wings \citep{1981ApJS...45..635V}.}
\tablenotetext{b}{Due to significant ISM absorption, the 1334.54 \AA~line was not included.}
\tablenotetext{c}{Average wavelength of the multiplet.}
\tablenotetext{d}{Fluxes uniformly corrected for 30$\pm$10\% ISM absorption (see Section \ref{sec:UVOpticalRelations} and \citealt{Youngblood2016TheExoplanets}).}
\end{deluxetable}

\begin{deluxetable}{cccccc}
\tabletypesize{\small}
\tablecolumns{8}
\tablewidth{0pt}
\tablecaption{Fit parameters for log$_{10}$(L$_{\text{H}\alpha}$/L$_{\text{Bol}}$) index and log$_{10}$ L$_{UV}$/L$_{BOL}$ \label{table:ha_correlations}} 
\tablehead{\colhead{Transition name} & 
                  \colhead{$\alpha$} &
                  \colhead{$\beta$} &
                  \colhead{$\rho$} & 
                  \colhead{$n$} &
                  \colhead{$\sigma$}
                  }
\startdata
SiIII & 0.99$\pm$0.07 & -1.66$\pm$0.32 & 0.78 & $<$0.0001 & 0.53 \\
LyA$^{a}$ & 0.58$\pm$0.08 & -1.41$\pm$0.35 & 0.74 & $<$0.0001 & 0.32 \\
NV & 1.02$\pm$0.11 & -1.44$\pm$0.48 & 0.78 & $<$0.0001 & 0.55 \\
SiII & 0.77$\pm$0.11 & -3.29$\pm$0.50 & 0.79 & $<$0.0001 & 0.42 \\
CII & 1.00$\pm$0.09 & -1.37$\pm$0.42 & 0.79 & $<$0.0001 & 0.58 \\
SiIV & 0.95$\pm$0.08 & -1.85$\pm$0.34 & 0.71 & $<$0.0001 & 0.60 \\
CIV & 0.91$\pm$0.07 & -1.38$\pm$0.32 & 0.82 & $<$0.0001 & 0.50 \\
HeII & 0.98$\pm$0.07 & -1.32$\pm$0.33 & 0.82 & $<$0.0001 & 0.49 \\
MgII*$^{,b}$ & 0.98$\pm$0.4 & 0.17$\pm$2.1 & 0.45 & 0.03 & 0.64 \\
\enddata
\tablecomments{The scaling relations take the form log$_{10}$ $L_{\rm UV}$ = ($\alpha$ $\times$ log$_{10}$(L$_{\text{H}\alpha}$/L$_{\text{Bol}}$)) + $\beta$, where $L_{\rm UV}$ represents each UV emission line luminosity in erg s$^{-1}$. $\rho$~is the Spearman correlation coefficient, $n$ is the probability of no correlation, and $\sigma$~is the standard deviation of the data points about the best-fit line (dex).}
\tablenotetext{*}{Fit with a double Gaussian because wings cannot be resolved at the lower signal to noise of these blended emission lines.}
\tablenotetext{a}{LP247-13 was not included in the fit because it is a significant outlier with a very small relative error on L(\Lya)/L$_{\rm Bol}$.}
\tablenotetext{b}{Fluxes uniformly corrected for 30$\pm$10\% ISM absorption (see Section \ref{sec:UVOpticalRelations} and \citealt{Youngblood2016TheExoplanets}). The fit does not include the significant outlier GJ676A. To account for scatter in the inactive range of the plot, we also fit a broken power law with a crossover point at log$_{10}$(L$_{\text{H}\alpha}$/L$_{\text{Bol}}$)=-5.0, separating the active and inactive regimes. The inactive regime is described by log$_{10}$ L$_{UV}$/L$_{BOL}$=-5.06$\pm$0.01, and the active regime by log$_{10}$ L$_{UV}$/L$_{BOL}$=(0.99$\pm$0.20)$\times$log$_{10}$(L$_{\text{H}\alpha}$/L$_{\text{Bol}}$)--0.14$\pm$0.82.}
\end{deluxetable}

\begin{deluxetable}{cccccc}
\tabletypesize{\small}
\tablecolumns{8}
\tablewidth{0pt}
\tablecaption{ Fit parameters for log$_{10}$ S and log$_{10}$ L$_{UV}$/L$_{BOL}$ \label{table:s_correlations}} 
\tablehead{\colhead{Transition name} & 
                  \colhead{$\alpha$} &
                  \colhead{$\beta$} &
                  \colhead{$\rho$} & 
                  \colhead{$n$} &
                  \colhead{$\sigma$}
                  }
\startdata
SiIII & 1.84$\pm$0.17 & -7.07$\pm$0.12 & 0.89 & $<$0.0001 & 0.39 \\
LyA & 1.05$\pm$0.19 & -4.53$\pm$0.13 & 0.81 & $<$0.0001 & 0.39 \\
NV & 1.73$\pm$0.17 & -6.96$\pm$0.12 & 0.88 & $<$0.0001 & 0.39 \\
SiII & 1.22$\pm$0.16 & -7.43$\pm$0.10 & 0.90 & $<$0.0001 & 0.31 \\
CII & 1.91$\pm$0.22 & -6.89$\pm$0.14 & 0.84 & $<$0.0001 & 0.59 \\
SiIV & 1.84$\pm$0.17 & -7.06$\pm$0.11 & 0.77 & $<$0.0001 & 0.61 \\
CIV & 1.87$\pm$0.16 & -6.45$\pm$0.11 & 0.91 & $<$0.0001 & 0.35 \\
HeII & 2.04$\pm$0.22 & -6.78$\pm$0.15 & 0.91 & $<$0.0001 & 0.41 \\
MgII*$^{,a}$ & 1.35$\pm$0.16 & -4.98$\pm$0.08 & 0.86 & $<$0.0001 & 0.37 \\
\enddata
\tablecomments{The scaling relations take the form log$_{10}$ $L_{\rm UV}$ = ($\alpha$ $\times$ log$_{10}$S) + $\beta$, where $L_{\rm UV}$ represents each UV emission line luminosity in erg s$^{-1}$. $\rho$~is the Spearman correlation coefficient, $n$ is the probability of no correlation, and $\sigma$~is the standard deviation of the data points about the best-fit line (dex).}
\tablenotetext{*}{Fit with a double Gaussian because wings cannot be resolved at the lower signal to noise of these blended emission lines.}
\tablenotetext{a}{Fluxes uniformly corrected for 30$\pm$10\% ISM absorption (see Section \ref{sec:UVOpticalRelations} and \citealt{Youngblood2016TheExoplanets}).}
\end{deluxetable}

\section{Discussion}\label{sec:Discussion}

Are our presented UV-optical scaling relations precise enough for photochemical and atmospheric escape models of exoplanets? In this section, we analyze the expected UV precision for a range of typical log$_{10}$\rhk{} values and measurement precisions. We focus on \rhk{} because those relations had the least scatter. Using a star with log$_{10}$\rhk{}$=-4.5\pm0.30$ ($\sim$7\% uncertainty) as an example, the precision of our UV luminosity estimates ($\sigma_L$/$L$) for each emission line ranges from factors of 2.27-4.65 (0.36-0.67 dex), depending on the specific emission line. We evaluated this precision by comparing the calculated error and the predicted average luminosity value. The uncertainties of each predicted UV emission line luminosity are dominated by the error of the best-fit line intercept and the error on log$_{10}$\rhk{}. From the standard deviations about the best fit lines (Table~\ref{table:rhk_correlations}), we estimate that using our scaling relations allows one to approximate the individual L$_{UV}$/L$_{bol}$ of the nine UV emission lines examined in this work within a factor of $\sim$2-4 (0.31-0.61 dex) for a typical M dwarf UV spectrum. Underscoring the utility and impact of these correlations is the fact that the parameter space for the L$_{UV}$/L$_{bol}$ of our target stars spans almost 3 orders of magnitude.

\cite{Rugheimer2015EFFECTSTARSb} examined the effect of variations in UV spectra on modeled exoplanet spectra and found that factor of $\lesssim$10 UV flux variations propagate to 10-30\% level changes in the depths of spectral features from simulated directly-imaged Earth-like planets. Depending on the precision of observed reflection spectra, using our \rhk{} scaling relations could be suitable for photochemical modeling purposes. For determining atmospheric escape rates from exoplanets, obtaining accurate EUV fluxes of M dwarfs is notoriously challenging due to a dearth of EUV spectra, and much of the exoplanet community relies on scaling relations between the EUV and other spectral regions like the FUV and X-ray \citep{Sanz-Forcada2011AstronomyEvaporation,Linsky2013THESTARS,Chadney2015XUV-drivenStars,France2018Far-UltravioletStars}. For some exoplanets needing atmospheric escape modeling, only an optical spectrum of the host star may be available, and here we estimate the suitability of our optical-FUV scaling relations for extrapolating to the EUV. \cite{Bolmont2017WaterTRAPPIST-1} showed that in the low- and high-EUV flux regimes, water loss rates on an Earth-like planet orbiting TRAPPIST-1 increase at the same rate as EUV flux. However, in the moderate EUV flux regime, water loss rates increase more slowly than a 1:1 relation with increasing EUV flux, indicating that uncertainties in the incident EUV flux up to a factor of 10 may be acceptable in this regime. However, determining where this moderate EUV regime is may depend on the particular star and simulated planet. Thus, it is unlikely that our optical-FUV scaling relations can be propagated into other FUV-EUV scaling relations (e.g., \citealt{Linsky2013THESTARS,France2018Far-UltravioletStars}) and retain a sufficiently small level of uncertainty that wouldn't dominate over the escape model's uncertainties. Based on these examples of models that consider the impact of absolute UV flux on exoplanets, we conclude that the precision provided by this work's scaling relations may be sufficient for photochemical modeling needs, but not atmospheric escape modeling. Further work is needed to demonstrate the impact of UV spectrum uncertainties on photochemical models, and will be addressed in an upcoming paper (Teal et al., in prep.).

\Lya{} alone represents 75-90\% of the 1200-1700 \AA~flux for typical M dwarfs (e.g., GJ 832, GJ 876, GJ 176) \citep{France2013THESTARS}, so by estimating (or directly measuring and reconstructing) the \Lya{} line, one can account for the majority of the FUV flux from an M dwarf. However, not accounting for the remaining spectral energy distribution across the FUV might significantly change results from photochemical models given the strong wavelength dependence of photoabsorption cross sections of key atmospheric molecules. Here we estimate the percentage of non-\Lya{}~FUV flux made up by the 7 FUV lines\footnote{Note that Mg II is a NUV doublet and is not included in the FUV analysis.} (excluding \Lya{}) we analyzed (Si II, Si III, Si IV, C II, C IV, He II, N V). An important limitation in our ability to characterize the FUV spectra of M dwarfs is the extremely faint FUV continuum (photospheric and chromospheric), which is well below the COS and STIS instrument background levels in almost all cases. \cite{Loyd2016ThePlanets} detected weak FUV continuum emission in 3/7 of the MUSCLES M dwarfs (GJ 832, GJ 876, GJ 176) by integrating across multiple line-free bandpasses and estimated that the continuum emission comprises at least 10\% of the 1307-1700 \AA~FUV flux region. Tilipman et al. in prep created high-resolution synthetic FUV spectra of GJ 832 and GJ 581 and found that the percentage of FUV emission between 1300-1700 \AA~comprised by continuum is 57\% and 43\%, respectively. Our ability to develop scaling relations for estimating the FUV continuum of M dwarfs depends on more sensitive observations of these stars as well as model stellar atmospheres that accurately treat the upper atmosphere \citep{Fontenla2016semi,Peacock2019PredictingSystem,Tilipman2020}. \cite{2012ApJ...745...25L} showed significant correlations between the FUV continuum of G dwarfs (measured over 1382–1392 \AA) and stellar rotation period as well as Si IV flux, so similar correlations likely exist for M dwarfs. 

By measuring the mean flux density in two line-free regions (1337-1351 \AA~and 1374-1392 \AA) from our spectra of GJ 832, GJ 876, and GJ 176, we estimate that weak FUV emission lines and continuum comprise 20-50\% of the non-\Lya{}~1200-1700 \AA~flux, while our 7 FUV lines (excluding \Lya{}) comprise 30-50\%. Other weak-to-moderate intensity emission lines that we did not include in our study comprise any remaining flux. This means that 50-70\% of the non-\Lya{}~FUV flux is unaccounted for by our scaling relations. A direct FUV spectrum may be best to accurately characterize the 1200-1210 + 1222-1700 \AA~FUV flux. However, if the weak forest of emission lines and FUV continuum is below the detector sensitivity, as is the case for essentially all but the brightest M dwarfs observable with HST, using the scaling relations to scale a high-S/N M dwarf spectrum could be an appropriate substitute for a direct FUV spectrum with HST. Similarly, our scaling relations do not account for flux from faint emission lines in the NUV or any chromospheric continuum. In the NUV, Mg II at 2796, 2802 \AA~is by far the brightest emission line, but there is a forest of much fainter Fe II lines as well as other atomic species that can be difficult to measure with HST depending on the brightness of the target. More work is needed to assess the impact of excluding faint but numerous emission lines from UV spectral inputs on photochemical models of exoplanet atmospheres.  

\begin{figure}
\centering
\includegraphics[width=0.7\textwidth]{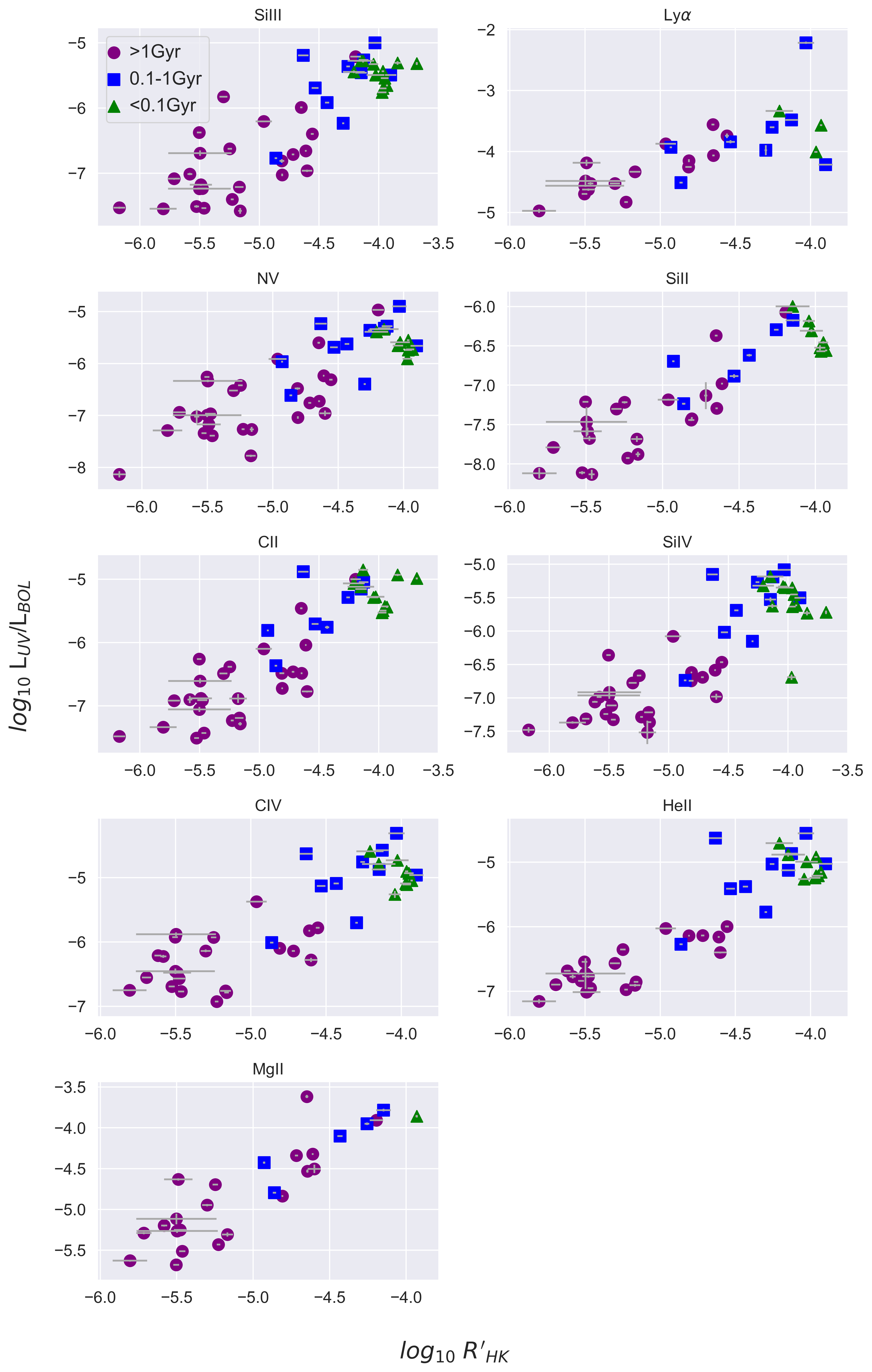}
\caption{Same as Figure \ref{rhk_vs_lum} with color break-down according to stellar age. Purple circles show the $>$1 Gyr (or field age) population, the blue squares show the 0.1-1 Gyr intermediate age population, and the green triangles show the young $<$0.1 Gyr ($<$100 Myr) population.}
\label{rhk_vs_lum_age_plots}
\end{figure}

\begin{figure}
\centering
\includegraphics[width=0.7\textwidth]{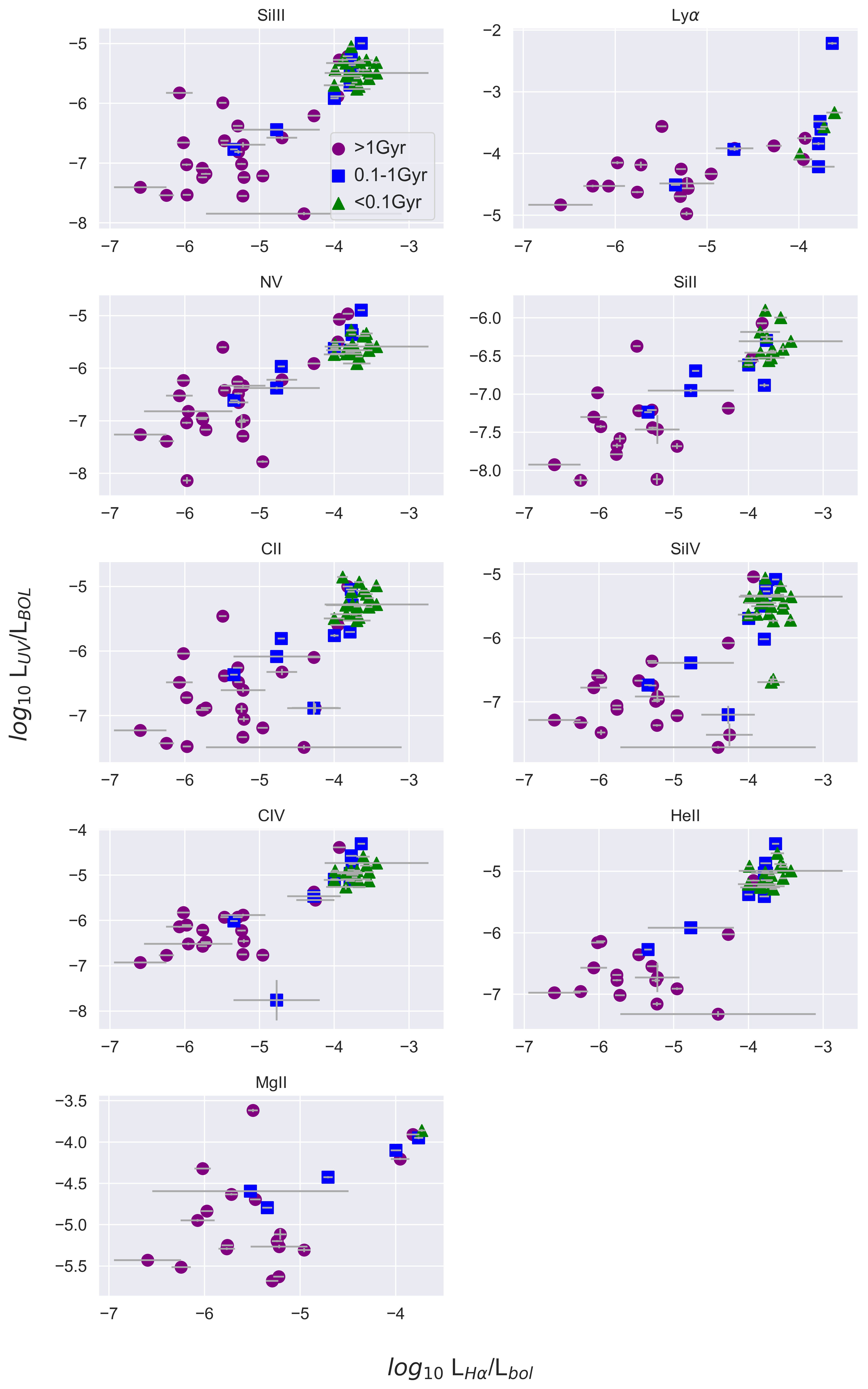}
\caption{Same as Figure \ref{Ha_vs_lum} with color break-down according to stellar age. Purple circles show the $>$1 Gyr (or field age) population, the blue squares show the 0.1-1 Gyr intermediate age population, and the green triangles show the young $<$0.1 Gyr ($<$100 Myr) population.}
\label{Ha_vs_lum_age_plots}
\end{figure}

\begin{figure}
\centering
\includegraphics[width=0.7\textwidth]{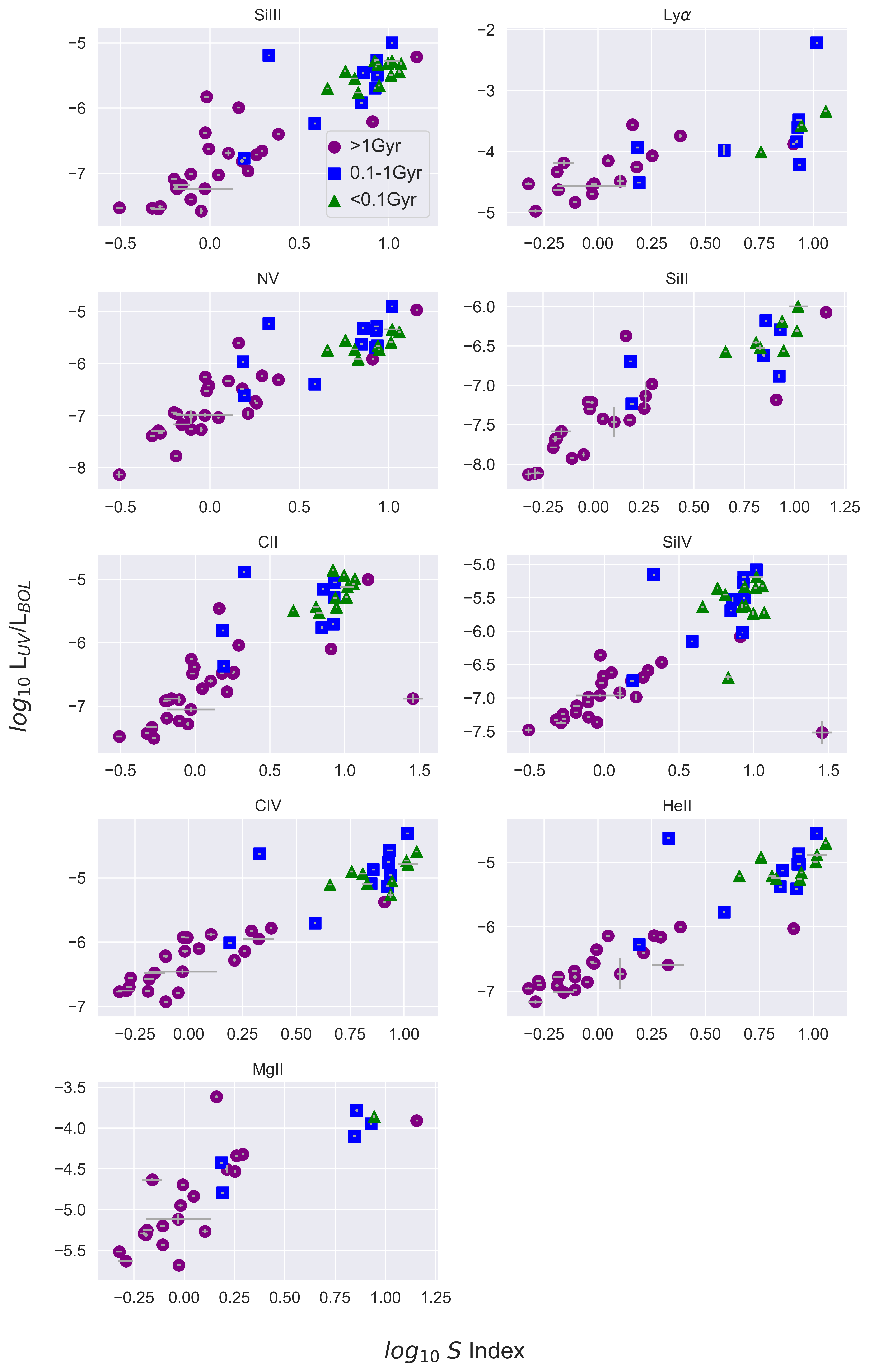}
\caption{Same as Figure \ref{s_vs_lum} with color break-down according to stellar age. Purple circles show the $>$1 Gyr (or field age) population, the blue squares show the 0.1-1 Gyr intermediate age population, and the green triangles show the young $<$0.1 Gyr ($<$100 Myr) population}
\label{s_vs_lum_age_plots}
\end{figure}

We also examine whether the scaling relations change based on the stellar age given that magnetic dynamos, which are ultimately responsible for the presence of these emission lines, change over a star's lifetime. We subdivided each plot of \rhk, S index, and \Lhabol{} as a function of UV activity into groups of stars $<$0.1, 0.1-1 Gyr (inclusive), and $>$1 Gyr as shown in Figures \ref{rhk_vs_lum_age_plots}-\ref{s_vs_lum_age_plots}. As expected, the oldest stars are typically the least active and the youngest are usually the most active. The youngest stars are clumped in the high activity region of the plots with no apparent linear relationship between optical and UV activity, whereas the intermediate and field age stars span a range of activity levels. We note that we have no stars in our sample with log$_{10}$ \rhk{} values $> -3.8$, which aligns with the finding in \cite{Astudillo-Defru2016MagneticRHK} that activity saturates around log$_{10}$\rhk{}$ = -3.5$. \cite{Stelzer2013TheSun}, \cite{Shkolnik2014HAZMAT.STARS}, and \cite{France2018Far-UltravioletStars} demonstrated a similar saturation limit with UV luminosity, although this limit varies depending on the emission line. We conclude that these young stars are saturated or nearly saturated in UV and Ca II luminosity, which explains the lack of correlation among them. However, we expect that a similarly narrow range of emission strength for older stars would also show a lack of correlation.

\section{Summary}\label{sec:Summary}
We have extended and improved upon previous efforts to determine useful scaling relations between the optical and UV spectra of M dwarfs. Through empirical analysis of four standard optical activity indices (\Ha{} equivalent width and \Lhabol, Ca II S index and \rhk), we have determined a new method of estimating the UV luminosity of M dwarfs when UV data is not available. The main findings are outlined below.

\begin{enumerate}
\item Time-averaged \rhk{}, S index, and \Lhabol{} correlate positively and significantly with the normalized UV luminosity (L$_{UV}$/L$_{bol}$) of nine far- and near-UV spectral lines (see Tables \ref{table:rhk_correlations}, \ref{table:ha_correlations}, and \ref{table:s_correlations}). The scatter about the best fit lines is lowest for \rhk{} (0.31-0.61 dex) and highest for the S index (0.58-0.84 dex). The scatter around the \Lhabol{} best fit lines ranges from 0.42-0.68 dex, excluding Mg II. No statistically significant correlation was found between L$_{UV}$/L$_{bol}$ and \Ha{} equivalent width. \label{no1} 
\item The luminosity of individual UV emission lines normalized to stellar bolometric luminosity can be estimated with \rhk{} within a factor of $\sim$2-4 (0.31-0.61 dex) (Table \ref{table:rhk_correlations}). This implies that the scaling relations defined in this study can be a useful substitute for direct UV observations of M dwarfs. \label{no4}
\end{enumerate}

\noindent The results presented here address important problems in the characterization of cool ($T_{eq} \lesssim 1000$ K) exoplanets and the search for habitable exoplanets. M dwarfs are excellent targets for finding and characterizing small and/or cool exoplanets. However, their UV spectra can have significant and misleading effects on the composition of exoplanet atmospheres. The UV-\rhk{} scaling relation developed in this paper provides an alternative method to completing photochemical analysis of exoplanet atmospheres without needing observations from space-based telescopes with valuable and limited resources. This will allow for efficient follow-up on exoplanet discoveries, which is essential given the number of current and upcoming dedicated exoplanet missions. In addition, this work will help determine which planets may be the most amenable for further study so that the outcome of observations on major missions like JWST can be maximized.

\acknowledgements

Based on observations with the NASA/ESA Hubble Space Telescope obtained from MAST at the Space Telescope Science Institute, which is operated by the Association of Universities for Research in Astronomy, Incorporated, under NASA contract NAS5-26555. Data used were obtained as parts of GO \#'s 13650, 14784, 14640, 13020, 14462, 14767, 12361, 11616, 12011, 9090, 15071, 15326, and 15190. This research also relied on the European Southern Observatory (ESO) Archive Facility for HARPS and UVES science products, and the Keck Observatory Archive (KOA) operated by the W. M. Keck Observatory for HIRES data. This work made use of spectral data retrieved from the ELODIE archive at Observatoire de Haute-Provence (OHP,  \url{http://atlas.obs-hp.fr/elodie/}) and is based in part on data products available at the Canadian Astronomy Data Centre (CADC) as part of the CFHT Data Archive as well. CADC is operated by the National Research Council of Canada with the sup- port of the Canadian Space Agency. This work was supported by a NASA Keck PI Data Award, administered by the NASA Exoplanet Science Institute. Data presented herein were obtained at the W. M. Keck Observatory from telescope time allocated to the National Aeronautics and Space Administration through the agency's scientific partnership with the California Institute of Technology and the University of California. The Observatory was made possible by the generous financial support of the W. M. Keck Foundation. The authors wish to recognize and acknowledge the very significant cultural role and reverence that the summit of Mauna Kea has always had within the indigenous Hawaiian community. We are most fortunate to have the opportunity to conduct observations from this mountain. This research also made use of the SIMBAD database, operated at CDS, Strasbourg, France. K.M. thanks Nicola Astudillo-Defru for helpful correspondence. She also acknowledges support from the NASA Internship program, the Universities Space Research Association (USRA) undergraduate scholarship awards, the National Space Grant Foundation's John Mather Nobel Scholarship Program to present this research, and the Bruce M. Babcock ‘62 Travel Research Fellowship to complete observations in Waimea, Hawaii. A.Y. and S.E.L. acknowledge support by an appointment to the NASA Postdoctoral Program at Goddard Space Flight Center, administered by USRA through a contract with NASA. R.O.P.L. and E.S. gratefully acknowledge support from NASA HST Grant HST-GO-14784.001-A for this work.

Data was graciously made available through the ESO archive from the following ESO programs: 072.C-0488(E), 183.C-0437(A), 198.C-0838(A), 077.C-0364(E), 191.C-0873(D), 191.C-0873(B), 191.C-0873(A), 082.C-0718(B), 183.C-0972(A), 085.C-0019(A), 091.C-0034(A), 090.C-0421(A), 191.C-0873(F), 191.C-0873(E), 095.C-0718(A), 192.C-0224(A), 191.C-0505(A), 192.C-0224(H), 192.C-0224(B), 089.C-0904(A), 095.D-0291(A), 088.C-0506(A), 095.C-0437(A), 082.C-0218(A), 180.C-0886(A), 093.C-0343(A), 076.C-0010(A), 074.C-0037(A), 60.A-9036(A), 192.C-0224(G), 192.C-0224(C), 096.C-0876(A), 097.C-0390(B), 099.C-0225(A), 68.D-0166(A), 075.C-0202(A), 099.C-0205(A), 075.C-0321(A), 082.D-0953(A), 099.C-0880(A), 096.C-0258(A), 089.C-0207(A), 077.C-0012(A), 079.C-0046(A), 080.D-0151(A), 276.C-5054(A), 086.D-0062(A), 081.D-0190(A), 089.C-0732(A), 093.C-0409(A), 095.C-0551(A), 096.C-0460(A), 092.C-0721(A), 192.C-0852(M), 098.C-0366(A), 088.C-0662(B), 089.C-0497(A), 076.C-0155(A), 495.L-0963(A), 074.B-0639(A).

\facilities{HST (COS, STIS), Keck:I (HIRES), ESO (HARPS, UVES), CFHT (ESPaDOnS), OHP (ELODIE)}

\software{ IPython \citep{Perez2007IPython:Computing}, Matplotlib \citep{Hunter2007Matplotlib:Environment}, Pandas \citep{Mckinney2010DataPython}, NumPy and SciPy \citep{VanDerWalt2011TheComputation}}

\bibliographystyle{aasjournal}

\bibliography{references.bib}
\newpage

\begin{figure}
\centering
\includegraphics[width=14cm]{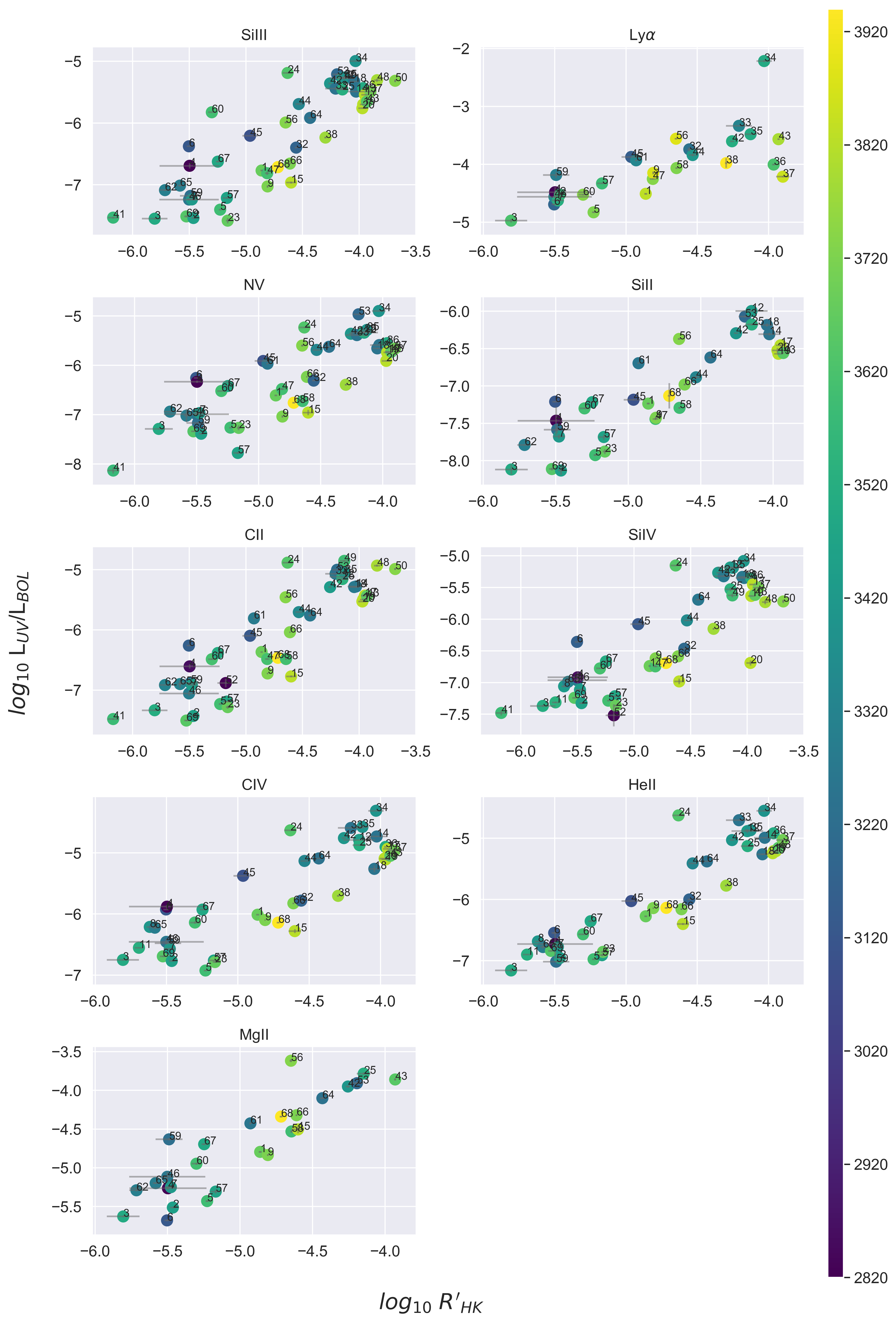}
\caption{Same graph as Figure \ref{rhk_vs_lum} with star names added. (Note: We plan on making Figures \ref{rhk_vs_lum}, \ref{Ha_vs_lum}, and \ref{s_vs_lum} interactive in the online journal. This figure and the two following are included so the referee can see the data point labels.))}
\label{LAB_rhk_vs_lum}
\end{figure}

\begin{figure}
\centering
\includegraphics[width=14cm]{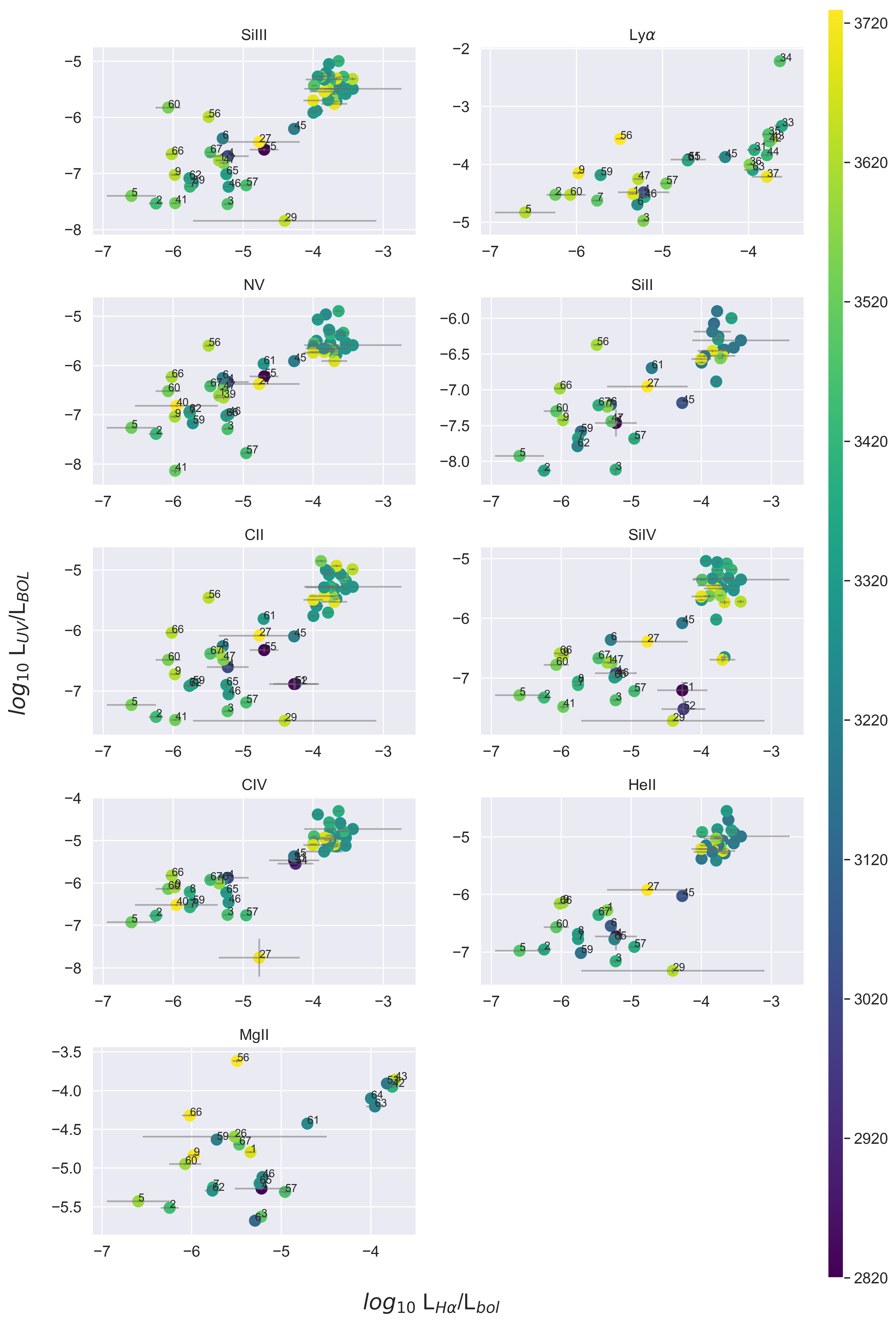}
\caption{Same graph as Figure \ref{Ha_vs_lum} with star names added.}
\label{LAB_Ha_vs_lum}
\end{figure}

\begin{figure}
\centering
\includegraphics[width=15cm]{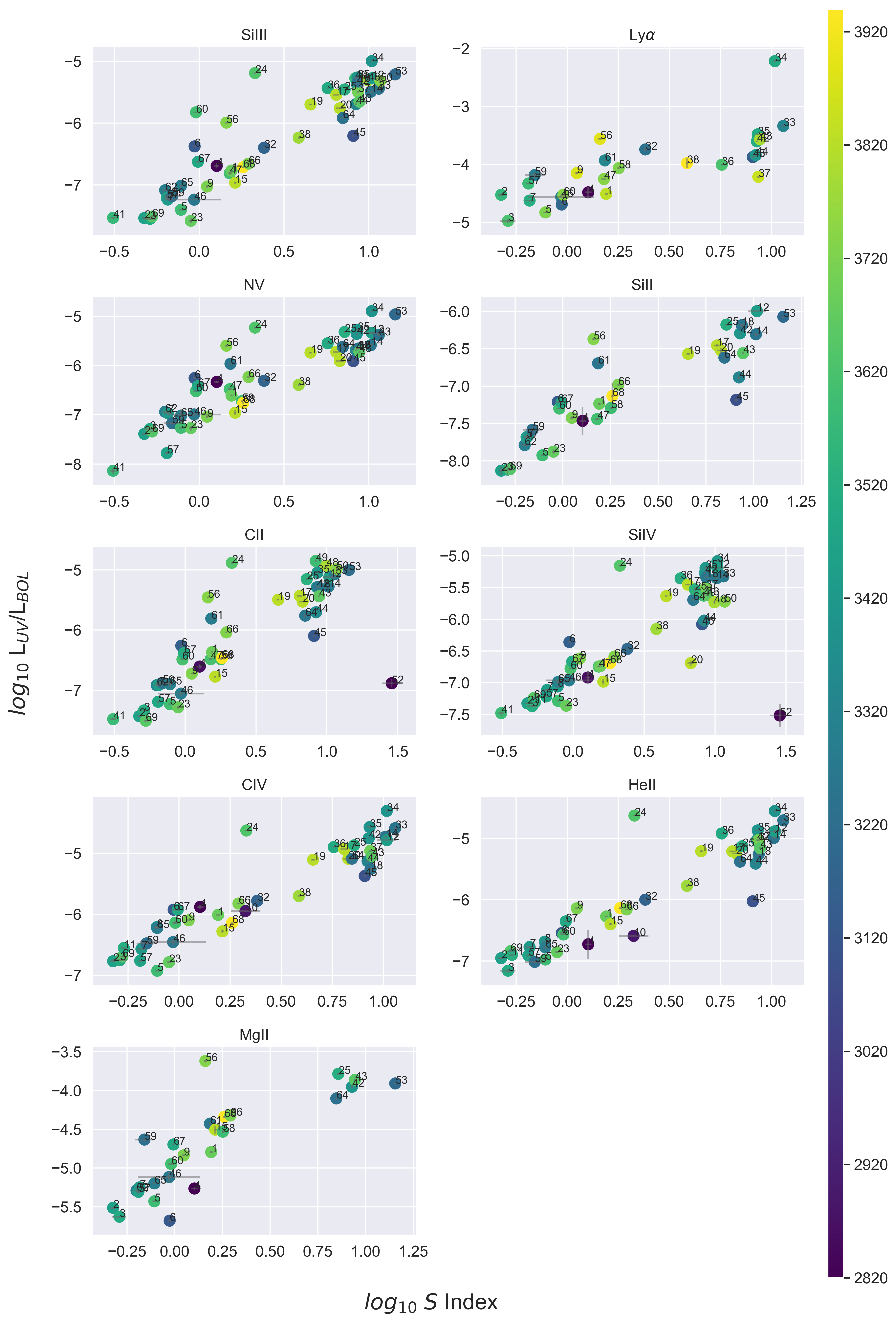}
\caption{Same graph as Figure \ref{s_vs_lum} with star names added.}
\label{LAB_s_vs_lum}
\end{figure}

\end{document}